\documentclass[a4paper,11pt]{article}
\usepackage{aaskaiid}
\usepackage{aas_macros}
\usepackage{orcidlink}

\title{Cosmological Galaxy Formation Modelling in the Era of the Square Kilometre Array}
\ShortTitle{Simulations in the SKA era}

\author[1]{Claudia del P. Lagos\orcidlink{0000-0003-3021-8564}}
\ShortName{Lagos et al.} % shortened name list for header 
\author[2]{Carlton M. Baugh\orcidlink{0000-0002-9935-9755}}
\author[1]{Connor Bottrell\orcidlink{0000-0003-4758-4501}}
\author[3]{Romeel Dave\orcidlink{0000-0003-2842-9434}}
\author[4]{Gabriella De Lucia\orcidlink{0000-0002-6220-9104}}
\author[3]{Jindra Gensior\orcidlink{ 0000-0001-6119-9883}}
\author[5]{Filip Hu{\v{s}}ko\orcidlink{0000-0002-1510-1731}}
\author[2]{Cedric G. Lacey\orcidlink{0000-0001-9016-5332}}
\author[1]{Danail Obreschkow\orcidlink{0000-0002-1527-0762}}
\author[2]{Kyle Oman\orcidlink{0000-0001-9857-7788}}
\author[1]{Chris Power\orcidlink{0000-0002-4003-0904}}
\author[1]{Nicole Thomas\orcidlink{0000-0003-4315-4555}}
\author[1]{Ruby J. Wright\orcidlink{0000-0002-1367-0949}}  
\author[6]{Lizhi Xie\orcidlink{0000-0003-3864-068X}}

\affiliation[1]{International Centre for Radio Astronomy Research (ICRAR), M468, University of Western Australia, 35 Stirling Hwy, Crawley, WA, 6009, Australia.}
\emailAdd{claudia.lagos@icrar.org}
\affiliation[2]{Department of Physics, Institute for Computational Cosmology, Science Laboratories, Durham University, South Road, Durham, DH1 3LE, UK.}
\affiliation[3]{Institute for Astronomy, Royal Observatory, University of Edinburgh, Edinburgh EH9 3HJ, UK.}
\affiliation[4]{INAF – Astronomical Observatory of Trieste, Via G. B. Tiepolo 11, 34143 Trieste, Italy.}
\affiliation[5]{Leiden
Observatory, Leiden University, PO Box 9513, 2300 RA Leiden, the Netherlands.}
\affiliation[6]{Tianjin Normal University, Binshuixidao 393, 300387 Tianjin, PR China.}

\abstract{
Over the past decade, galaxy formation simulations have advanced dramatically, transforming our ability to model the interstellar medium (ISM) and predict galaxies’ radio emission. Yet the challenge of bridging physical scales—from sub-parsec star formation to gigaparsec cosmic structure—remains. The Square Kilometre Array (SKA) will map the cold gas and radio continuum of galaxies across cosmic time, demanding models that couple physical realism with cosmological reach.
This chapter reviews the state-of-the-art in cosmological galaxy formation modelling in preparation for the SKA. We outline progress in simulating atomic hydrogen (HI), molecular gas, and radio continuum emission from both star formation and active galactic nuclei, highlighting how cosmological hydrodynamical simulations and semi-analytic models now jointly reproduce many observed gas properties. We emphasise the need for a coordinated, ``wedding-cake'' strategy that unites simulations of different scales, for forward modelling of observables to ensure fair comparison with data, and for the integration of new technologies such as AI-driven emulators to accelerate progress. Together, these efforts will enable theoretical models to both interpret and guide SKA science, turning simulations from passive interpreters into active engines for discovery.
}
\begin{document}
\maketitle

\section{Introduction}

A variety of numerical techniques exist to simulate galaxy formation and evolution, each with distinct advantages and limitations. To understand these differences, it is essential to establish the relevant physical scales involved in the problem of galaxy formation and to identify which processes are directly resolved in simulations versus those that must be approximated through so-called subgrid models. These models encapsulate the macroscopic effects of unresolved physics that occur below the resolution limit of a simulation.

On the largest scales, the cosmic web and the evolution of density fluctuations are typically directly simulated in cosmological models. Similarly, on megaparsec scales, the internal structure of dark matter haloes, such as their density and velocity profiles, is also generally well resolved. Within haloes, many baryonic processes, including gas accretion and large-scale magnetohydrodynamic flows, can be directly followed using modern hydrodynamical codes \citep{Springel.2010,Hopkins.2017,Schaller.2024} that can also incorporate radiative transfer. However, several critical processes, such as metal-line cooling of gas, are often still implemented via subgrid prescriptions.

At smaller, galactic scales, subgrid modelling becomes essential. The physics of the interstellar medium (ISM), including dust shielding, gas cooling, and the formation of a multiphase ISM, is typically treated through phenomenological models. Processes occurring on even smaller scales, such as star formation, stellar evolution, and the growth and feedback of supermassive black holes, are entirely subgrid in nature.

An important aspect of galaxy formation theory is that these physical scales are inherently coupled: large-scale processes influence the conditions within galaxies, while small-scale processes, such as feedback, can propagate effects to cosmological environments (e.g. \citealt{Hellwing.2016, Tillman.2023, Gebhardt.2024, Schaller.2025}), making astrophysics and cosmology intimately intertwined. The transition between directly simulated and subgrid-modelled physics varies across simulation suites, depending on their resolution and numerical approach. Ideally, simulations aim to resolve as much physics as possible to maximise physical insight, but computational constraints invariably necessitate subgrid treatments.
Subgrid models typically include free parameters that control efficiencies of various processes. These parameters are often calibrated to reproduce selected observables.
%, an essential but sometimes under-appreciated aspect of galaxy formation modelling that influences predictive power and interpretation.

\begin{figure}
    \centering
    \includegraphics[width=0.99\linewidth]{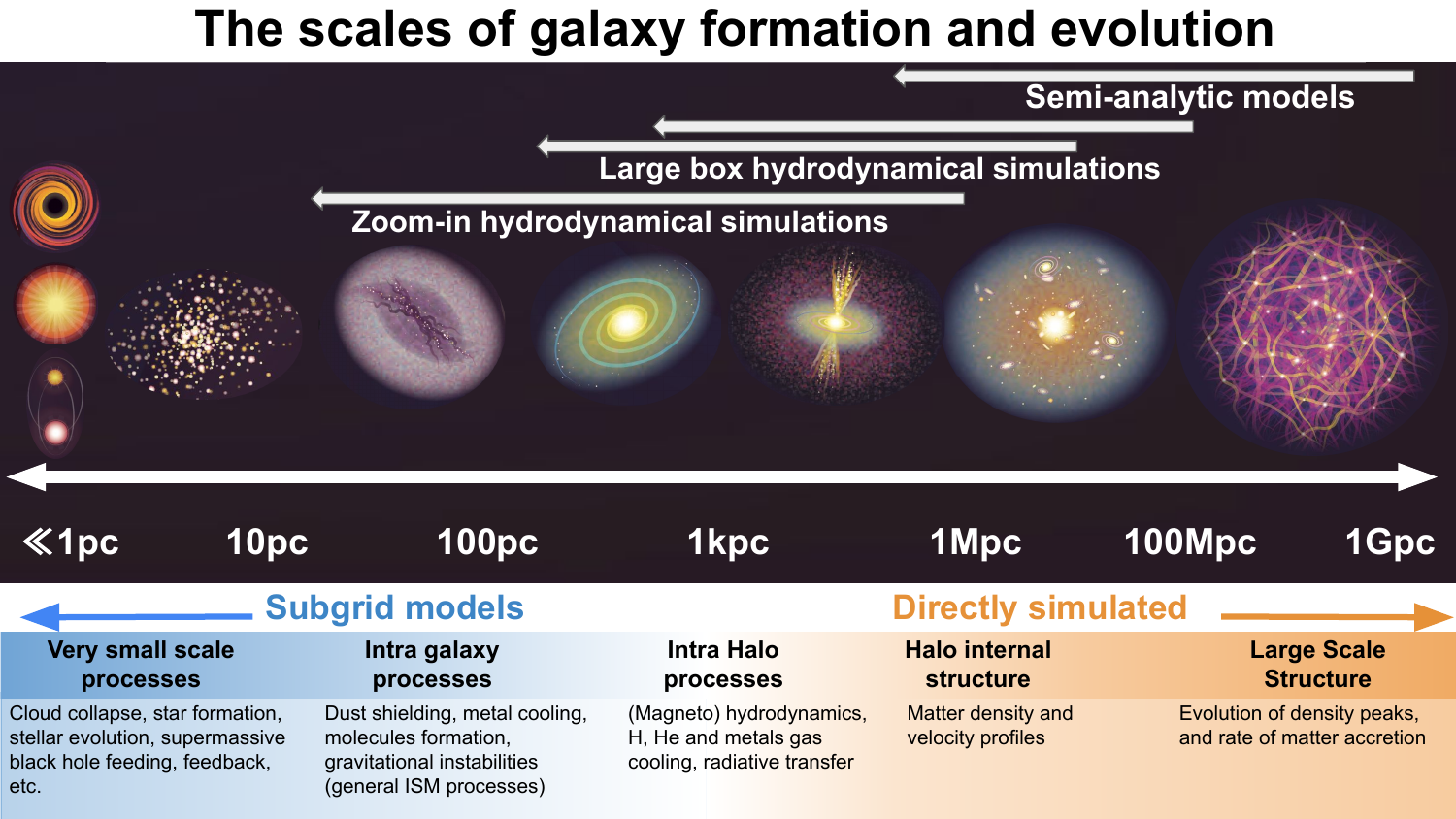}
    \caption{Modified from \citet{Crain.2023} to show the processes that are happening at different scales and place them into the context of what is directly simulated and what is modelled via so called ``subgrid physics models''. The exact transition of when one goes from directly simulating to modelling depends on the tools at hand. The arrows at the top show the regions that are typically directly simulated in different techniques based on the latest models and simulations.}
    \label{fig:scales}
\end{figure}

In this context, semi-analytic models (SAMs) and cosmological hydrodynamical simulations occupy complementary regions of parameter space. SAMs can efficiently explore large cosmological volumes, albeit by invoking a greater number of subgrid prescriptions and approximations to describe halo and galactic processes. Hydrodynamical simulations, in contrast, can resolve smaller physical scales more explicitly but are limited in the cosmological volume they can represent. Large simulation programmes often adopt a ``wedding-cake'' strategy, employing a hierarchy of simulation boxes of varying size and resolution to bridge these regimes (e.g. \citealt{Pillepich.2018, Schaye.2015, Schaye.2025}). To reach resolutions of 1-10 parsecs, one must use zoom-in simulations, which focus on full hydrodynamical simulations of small cosmological volumes, typically chosen from large N-body boxes. These simulations can provide a more detailed physical view of the processes internal to galaxies, but usually are limited to a handful of objects \citep{Hopkins.2023,Wang.2015,Agertz.2020,Reina-Campos.2022}. 
Fig.~\ref{fig:scales} shows a schematic (albeit very simplified) view of the problem, also placing into context the three tools mentioned above to study galaxy formation and evolution: SAMs, large box cosmological hydrodynamical simulations and zoom-in simulations. The positions of the arrows are only indicative with the objective of creating a sense of where these different tools may be most powerful. 
Ultimately, subgrid models remain a cornerstone of galaxy evolution simulations. They encapsulate the unresolved physics that underpins much of the emergent phenomenology, yet their influence is often under-discussed. Understanding their role in the predicted properties of galaxies is thus critical.

In most state-of-the-art cosmological simulations and models, the free parameters of subgrid prescriptions are calibrated primarily to match the $z = 0$ stellar mass function (e.g. \citealt{Crain.2015,Chaikin.2025,Pillepich.2018,Lagos.2024,DeLucia.2024}). Although this approach ensures broad consistency with the local galaxy population, it inevitably introduces degeneracies in the underlying physical solutions. 
\citet{Mitchell.2018} carried out a direct comparison between the {\sc Eagle} hydrodynamical simulation and the GALFORM semi-analytic model. Despite employing very different methodologies, both models produced very similar stellar mass growth histories. However, when examining the gas reservoirs, striking discrepancies emerged: the predicted ISM masses differed by factors of 2–3, and the circumgalactic gas by factors of 5–6. \citet{Wright.2024} compared the baryon cycle in three widely-known hydrodynamical simulation projects: {\sc Eagle} \citep{Schaye.2025}, Illustris-TNG (a.k.a. TNG, \citealt{Pillepich.2018}), and Simba \citep{Dave.2019}. Because these models are tuned to the same $z = 0$ stellar mass function, they naturally produce comparable stellar-to-halo mass relations, with variations below 50\%. Yet, their predicted baryon contents differed by factors of up to five, displaying complex, model-dependent trends with halo mass. These variations originate from fundamental differences in the way in which simulations implement feedback and gas flows.
Together, these comparisons demonstrate that gas, in its multiple phases and reservoirs, provides a set of constraints that are largely orthogonal to those obtained from stellar masses or star formation rates. Gas traces the flow of baryons into, within, and out of galaxies, directly probing the physical mechanisms that regulate star formation and feedback—arguably the least understood components of galaxy formation theory.

This is precisely where the Square Kilometre Array (SKA) will play a transformative role. By mapping the distribution and dynamics of neutral hydrogen across cosmic time, the SKA will provide statistically powerful measurements of cold gas reservoirs in galaxies and their environments across cosmic epochs. These data will offer the missing empirical anchor for constraining feedback efficiency and baryon cycling in simulations, breaking long-standing degeneracies among competing galaxy formation models. In this way, the SKA will supply the critical observational leverage needed to disentangle how galaxies acquire, retain, and expel their gas—the processes that ultimately drive their formation and evolution. 
In this chapter we discuss how cool gas, and in particular atomic hydrogen, is currently modelled in galaxy formation models and simulations, and what are key predictions arising for the distribution of HI internal and external to galaxies, as well as how this evolves over time. We also touch upon what simulations are predicting for radio continuum and how that can be used to constrain the most uncertain mechanisms invoked in galaxy formation simulations. 

This chapter is organised as follows Section~\ref{sec:modelling} describes the current state-of-the-art in the modelling of HI and radio continuum emission, describing the process in very high resolution simulations, large-box cosmological hydrodynamical simulations to SAMs. Section~\ref{sec:HIprogress} focuses on recent progress and challenges modelling the HI emission of galaxies focusing on how HI can be used a cosmological probe and as a probe of subgrid modelling. Section~\ref{sec:contprogress} introduces recent progress in the modelling of radio continuum emission associated with star formation and AGN in galaxies, focusing on recent predictions that can be tested with the SKA. Section~\ref{sec:future} discusses how all the methods introduced in Section~\ref{sec:modelling} can be used in tandem to get the best of SKA observations, also focusing on what areas need dedicated effort in the coming years to better combine different galaxy evolution methods. Finally, Section~\ref{sec:summary} summarises our main findings.

\section{Modelling radio sources over all relevant scales}\label{sec:modelling}
%atomic hydrogen over all relevant scales}

In this section, we briefly introduce the methods being used by simulations to model the HI content of galaxies and the radio continuum emission, covering most of the extragalactic science priorities of the SKA.

\subsection{Very high resolution simulations}

\noindent {\it Modelling Atomic Hydrogen.} In recent years, there has been an increasing number of cosmological-style simulations that model gas cooling down to temperatures of $\lesssim$10 K. This signifies an important improvement over the status quo a decade ago of imposing a much higher temperature floor ($\approx 10^4 \,\rm K$), which prevented simulations from carrying out more detailed ISM studies. The reasoning behind the adopted temperature floor or ISM equation of state was to avoid artificial fragmentation due to limited resolution, but this simplification suppressed realistic multiphase ISM structure and altered the interplay between cooling and feedback. There has been a major shift in the last years from this modelling assumption leading to a new suite of (mostly) cosmological zoom-in simulations (e.g. MUGS, \citealt{Stinson.2010,Keller.2016}; FIRE, \citealt{Hopkins.2014,Hopkins.2018,Hopkins.2023}; NIHAO, \citealt{Wang.2015,Blank.2019}; EDGE, \citealt{Agertz.2020}; EMP-\emph{Pathfinder}, \citealt{Reina-Campos.2022}) or cosmological volumes (e.g. FIREbox, \citealt{Feldmann.2023}; {\sc Colibre}, \citealt{Schaye.2025}) to have a model of the ISM that self-consistently includes a cold and dense gas phase. 

In addition to more detailed ISM physics, these simulations (excepting the larger {\sc Colibre} volumes) tend to have a relatively high mass resolution of $\lesssim 3\times 10^5$ $\rm{M}_{\odot}$, and gravitational softening lengths ranging from a few to a couple of hundred parsec. Whilst there are some differences in how the neutral Hydrogen fraction is calculated at runtime (i.e. whether a non-equilibrium chemistry solver is employed, if it also includes molecular hydrogen as a separate species or whether this contribution has to be removed in post-processing, and full radiative-transfer limited to very high-resolution simulations of dwarf galaxies, e.g. EDGE), this allows the simulations to resolve the neutral gas within galaxies on small scales. As the structure of HI within the disc of a galaxy is set by the interplay of thermo-chemistry, star formation and feedback processes across cosmic time, this can be used to study the impact of different baryonic physics modelling choices. For example, FIREbox and EMP-\emph{Pathfinder} with a turbulence-dependent star formation efficiency can reproduce the thin HI discs and their characteristic flaring at larger radii observed in the local Universe \citep[see e.g.][]{Gensior.2023,Gensior.2024}. Different models of stellar feedback affect the turbulent structure of the gas \citep[e.g. reflected in the power spectrum:][]{Walker.2014}. Different density thresholds imposed on the star formation model can lead to different numbers of features identified in HI surface density maps \citep[e.g.][]{Maccio.2022}. Similarly, \citet{Gensior.2024} found that the HI disc morphology is the best predictor for the underlying physics with which galaxies were simulated (comparing galaxies from EMP-\emph{Pathfinder} simulated with different star formation models, and FIREbox). Differences in the HI morphology, driven by differences in physics models, seem to persist also on larger scales (e.g. \citealt{Marasco.2025}, see also Section 3). %Extending these small-scale predictions to more simulation suites with different physics models, and subsequently comparing them to a large sample of similarly-resolved observed galaxies has the potential to disprove or validate some of the underlying physics models. 

{\it Modelling radio continuum.} For radio continuum emission, non-cosmological high resolution simulations have been key to understand the complexity of the radio emission, especially coming from AGN-driven jets (e.g. \citealt{Kramer.2024,Jerrim.2025}). These simulations tend to be compared with individual AGN observations, such as those focusing on Centaurus A, and have been essential to demonstrate that allowing for explicit radiative losses in the models are essential for producing realistic jet brightness. However, bridging those very high-resolution simulations with cosmological ones has been challenging, and communities remain largely separated. Attempts to connect them have been done using more approximate modelling which we describe in the next two sections. 

\subsection{Large box cosmological hydrodynamical simulations}\label{sec:cosmosims}
%Romeel, Claudia, Kyle, etc  [current text from Romeel]

Large-volume galaxy formation simulations track the state of the gas dynamically, subject to hydrodynamic forces, radiative processes, and galactic feedback. While mostly lacking in resolution and therefore relying on a subgrid treatment of the ISM compared with models in the previous section, they provide good statistics sampling a wide range of environments.

\noindent {\it Modelling Atomic Hydrogen.} Early simulations computed HI in post-processing such as \citet{Dave.2013}, Illustris \citep{Vogelsberger.2014}, {\sc Eagle} \citep{Lagos.2015,Crain.2017} and TNG \citep{Stevens.2019} by assuming a self-shielding correction to separate HI from ionized hydrogen, and another prescription for separating out molecular from atomic hydrogen. Simba \citep{Dave.2019} computed HI fractions on the fly, but still used prescriptions that mimicked the post-processing approaches. The map-based approach to calculating HI fractions in post-processing (e.g. \citealt{Diemer.2018,Stevens.2019}) has some strengths, but immediately precludes working with kinematics. For example, it is impossible to construct a data cube because you can only know the integrated column density at each position. This is further motivation for the importance of self-consistent on-the-fly HI fractions from simulations. 

\begin{figure}
    \centering
    \includegraphics[trim=0mm 2mm 2mm 2mm, clip,width=0.99\linewidth]{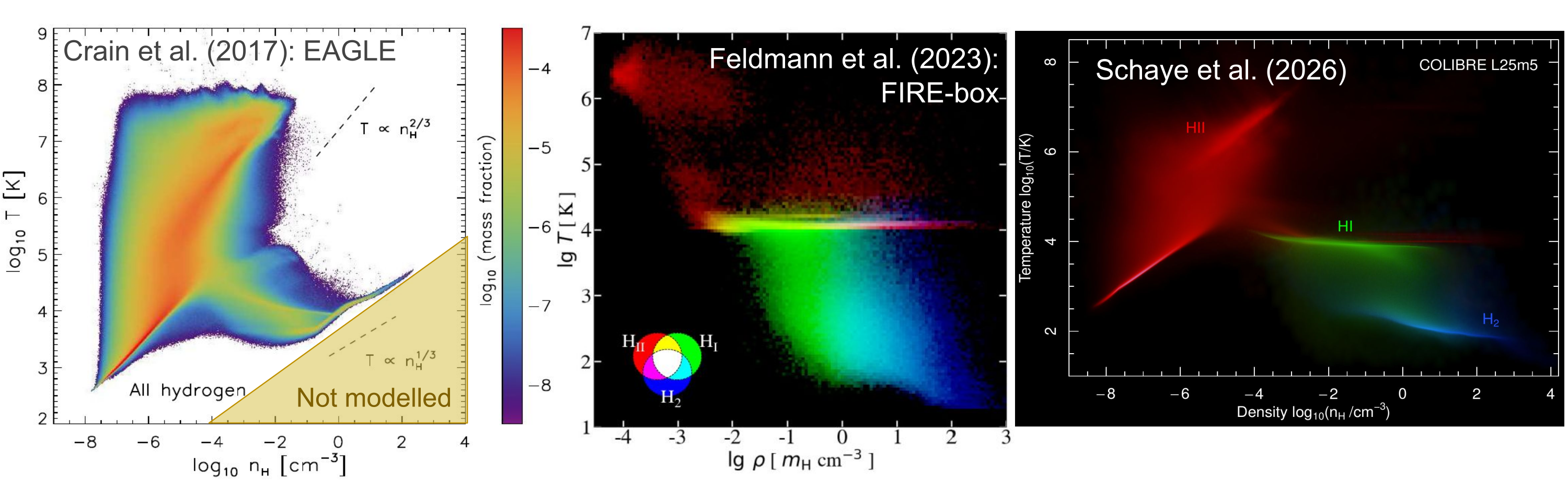}
    \caption{The distribution of gas particles or cells in the temperature vs density plane at $z=0$ in three different cosmological hydrodynamical simulations. {\sc Eagle} (left panel) \citep{Crain.2017} lacked an explicit model of the ISM and instead imposed an equation of state. Here, pixels are coloured by the mass fraction, showing that most of the gas mass in the $z=0$ universe is in the form of hot, low-density gas. FIREbox (middle panel) \citep{Feldmann.2023} uses cooling tables to allow the gas to cool down to $10$~K and account for the HII, HI, H$_2$ partition in post-processing. Here, pixels are coloured by the fraction of HII (red), HI (green) and H$_2$ (blue) mass. {\sc Colibre} (right panel) \citep{Schaye.2025} includes explicit on-the-fly non-equilibrium chemistry and cooling that explicitly tracks the HII, HI and H$_2$ species. Here, pixels are coloured by the fraction of HII (blue), HI (red) and H$_2$ (green) mass. The middle and right panel are examples of the recent progress in the field.}
    \label{fig:phasespace}
\end{figure}

Recent simulations focus on exactly that, including {\sc NewHorizon} \citep{Dubois.2021}, {\sc FIREbox} \citep{Feldmann.2023}, and {\sc Colibre} \citep{Schaye.2025}. These simulations marry a subgrid ISM model with a large volume to yield HI fractions more self-consistently over a large sample. As HI observations improve, simulations are commensurately focusing more on accurate predictions for HI. Fig.~\ref{fig:phasespace} shows examples of the distribution of particles in the {\sc Eagle}, FIREbox and {\sc Colibre} simulations, to show the parameter space that has been unlocked by the latest simulations: cold and high density gas, typical of the ISM. We note that enabling such sophisticated ISM models is not at all trivial, and has happened in steps over the last decade. We highlight that \citet{Richings.2022, Ploeckinger.2024} showed the importance of such modelling by demonstrating the large impact it could have on the HI and H$_2$ content of gas in different phases, and motivated the need for this modelling in cosmological settings. These simulations can start investigating the emergence of the Kennicutt-Schmidt \citep{Kennicutt.1998} relation for the atomic and molecular gas independently \citep{Kraljic.2024,Lagos.2025b}.
An important consideration is that cosmological hydrodynamical simulations with sufficient resolution to probe the HI content of galaxies down to $10^9\,\rm M_{\odot}$ typically have volumes of $\lesssim (100\,\rm cMpc)^3$ (with cMpc referring to comoving mega parsec). Larger volume simulations are possible but at lower resolution (e.g. \citealt{Schaye.2023,Schaye.2025}), which becomes prohibiting for the study of the gas content of galaxies.

As stated in the introduction, most simulations are calibrated to match stellar properties such as the galaxy stellar mass function.  Gas properties, on the other hand, provide unconstrained tests of the models. \citet{Dave.2020} compared Simba, {\sc Eagle}, and TNG versus local Universe observations of HI and H$_2$ in galaxies. The differences were substantial, particularly on their predicted evolution (see Section~3.3). For instance, Simba predicted an increasing HI mass function out to $z\approx 1$, while TNG's decreased (we will come back to this in Section~3.3). Precursor surveys such as LADUMA on MeerKAT and GMRT surveys \citep[e.g.][]{Chowdhury.2024} can provide preliminary constraints via stacking or a handful of distant 21cm-bright galaxies, but SKA's sensitivity will be crucial for providing definitive answers.

Ancillary data for 21cm-detected galaxies will be critical for maximizing the constraining power of SKA data \citep{Duncan01.2026.SKA}. Combining with photometric information opens up comparisons to quantities such as HI scaling relations \citep{Dave.2020}, the baryonic Tully-Fisher relation \citep{Glowacki.2020}, the angular momentum--mass relation \citep{Elson.2023,Hardwick.2023}, and HI asymmetries \citep{Watts.2020,Glowacki.2022}.  Conversely, 21cm surveys targeting a specific photometric or (preferably) stellar mass limits offer opportunities to constrain simulations via stacking as well as to explore the transition to HI-poor galaxies particularly as a function of environment (e.g. \citealt{Roychowdhury.2022}). Currently many of these comparisons are limited by the depth and/or resolution of HI data, motivating surveys with the SKA to fully unleash the power of HI to constrain models of galaxy evolution. 

\noindent {\it Modelling radio continuum.} Most of the efforts to predict radio continuum emission in large box cosmological hydrodynamical simulations have focused on AGN (e.g. \citealt{Thomas.2025,Husko.2025}) and the emission associated with the production of jets. \citet{Thomas.2025} estimated the radio continuum emission associated with galaxies and AGN by applying simple scaling relations between the star formation rate and the black hole accretion rate with the expected $1.4$~GHz emission. Although the method is potentially too simplistic, it already opens the possibility to compare with the latest data coming from SKA pathfinders and test the astrophysics models implemented in the simulations.  
%This was used to compare with current radio continuum observations 

In Sections 3 and 4 we will be showing results from TNG  (using the 100~Mpc and 50~Mpc boxes; \citealt{Pillepich.2018,Pillepich.2019}); {\sc Eagle} \citep{Schaye.2015}; Simba \citep{Dave.2019}; {\sc Colibre} \citep{Schaye.2025}; and FIREbox \citep{Feldmann.2023}. Of these runs, TNG, {\sc Eagle} and Simba do not model the cold ISM of galaxies, while {\sc Colibre} and FIREbox do.

\subsection{Galaxy formation semi-analytic models}
%Claudia, Gabriella, Lizhi, Cedric, Danail, etc  [text from Lizhi]

\noindent {\it Modelling Atomic Hydrogen.} Semi-analytic models provide a flexible approach for simulating a large cosmological volume at low computational cost, as they are built upon the skeleton constructed by dark matter only simulations. Semi-analytic models treat cold gas as a discrete component and for the most part do not account for its spatial distribution, temperature, or dynamics (albeit see \citealt{Stevens.2019b}). Current semi-analytic galaxy formation models self-consistently calculate the HI and H$_2$ fractions alongside  H$_2$-based star formation \citep{Fu.2010, Lagos.2011a, Somerville.2015, Stevens.2016, Xie.2017, Lagos.2018}, by implementing sub-grid physics based on scaling relations between the gas surface density, chemical abundances, mid-plane pressure, and star formation rate. These relations are drawn from either empirical \citep{Blitz.2006}, theoretical \citep{Krumholz.2009b}, or high-resolution simulation studies \citep{Gnedin.2011,Gnedin.2014} and are benchmarked to sub-kpc-scale detections of nearby galaxies. Many independent studies show that HI fractions, while insensitive to the HI-H$_2$ partition method, are primarily determined by stellar mass and gas disc size, and are significantly influenced by environmental effects. Although the recipes used in semi-analytic models are simple and heavily based on local Universe small-scale scaling relations, they appear to lead to predictions that are broadly consistent with the evolution of the cold gas content of galaxies (e.g. \citealt{Lagos.2014}), which is quite remarkable. This suggests that on small scales, the relationship between gas surface density, star formation rate and pressure or metal content is not strongly evolving, at least out to $z\approx 2$.

Besides the cosmological volumes probed by semi-analytic models (typically $\gtrsim (500\,\rm cMpc)^3$) compared to cosmological hydrodynamical simulations, there is a more subtle advantage to semi-analytic models, which is that the HI content of galaxies converges down to much lower masses than in cosmological hydrodynamical simulations, allowing them to probe a much larger dynamic range (see \citealt{Wen.2025} for a recent example). This is not surprising given the methodology employed, but it is important given that 21-cm surveys are sensitive to gas-rich galaxies, many of which are dwarf galaxies \citep{Maddox.2015}.

\noindent {\it Modelling radio continuum.} There has been significant recent progress in the modelling of radio continuum associated with star formation and AGN. The most recent example is \citet{Hansen.2024}, which presented a physically motivated model for the radio continuum within the {\sc Shark} \citep{Lagos.2018, Lagos.2024}. The emission mechanisms associated with star formation included radio emission split into free–free and synchrotron components. Free–free emission is linked to the density of free electrons in HII regions surrounding massive young stars, while synchrotron emission is tied to core-collapse supernova rates (the main sources of cosmic-ray electrons), with a small additional contribution from supernova remnants. In the case of AGN, \citet{Hansen.2024} extended  the \citet{Fanidakis.2011} framework, linking jet power to the black hole’s mass, accretion rate, and spin, and explicitly including synchrotron self-absorption component. The model distinguishes between thin-disc and advection-dominated accretion flow (ADAF) regimes, capturing the transition from radiatively efficient to inefficient modes. The introduced framework in \citet{Hansen.2024} for radio continuum emission allows the models to move from empirical calibrations to physically motivated, self-consistent frameworks within galaxy formation models.

In sections 3 and 4, we will be showing results from three semi-analytic models as examples. They are the GALFORM \citep{Lacey.2016}, GAEA \citep{DeLucia.2024} and {\sc Shark} \citep{Lagos.2018,Lagos.2024} semi-analytic models. These models are characterised by large volumes (typically $(500-800\,\rm Mpc)^3$) and apply similar ISM models, in which stars form from H$_2$ gas. They, however, can differ in their treatment of other subgrid models, such as AGN feedback (see \citealt{Lagos.2025} for a description of these three models and a comparison).

\section{Recent progress and challenges in modelling atomic hydrogen}\label{sec:HIprogress}

\subsection{HI kinematics}\label{sec:kinematics}
%Jindra, Kyle, Connor

Semi-analytic galaxy formation models can struggle to capture even basic kinematic properties of HI in galaxies, such as the spectral line width distribution \citep[e.g.][but see \citealp{Chauhan.2019}]{Brooks.2023}; hydrodynamical simulations are much better suited for detailed studies of the internal kinematics of galaxies. While these are able to capture many aspects of the complex, multifaceted physics driving HI kinematics -- including galaxy interactions, star formation \& feedback, disc \& hydrodynamical instabilities, etc. -- they are still limited. Perhaps the most severe limitation is that the atomic gas phase usually coexists with other phases within a single multi-phase resolution element (particle or gas cell). This imposes an overly strong coupling between the thermodynamics and kinematics of gas in different phases. As an example, HI in models with a temperature floor near $10^4\,\mathrm{K}$ (e.g. {\sc Eagle}, TNG, Simba) has temperatures, densities, pressures and velocity dispersions typical of such warm gas \citep{Crain.2017,Diemer.2018,Ploeckinger.2025}. While this may adequately capture the warm neutral medium (WNM) in galaxies, the cold neutral medium (CNM) is very poorly represented. While gas can be tagged in post-processing as best conceptualised as CNM gas, its kinematics inevitably remain artificially warm. This is usually most immediately evident as elevated velocity dispersions, often of $20$-$30\,\mathrm{km}\,\mathrm{s}^{-1}$ for HI, compared to the $\sim 12\,\mathrm{km}\,\mathrm{s}^{-1}$ typically observed \citep[e.g.][]{Pillepich.2019,Jimenez.2023}.

\begin{figure}
    \centering
    \includegraphics[width=0.99\linewidth]{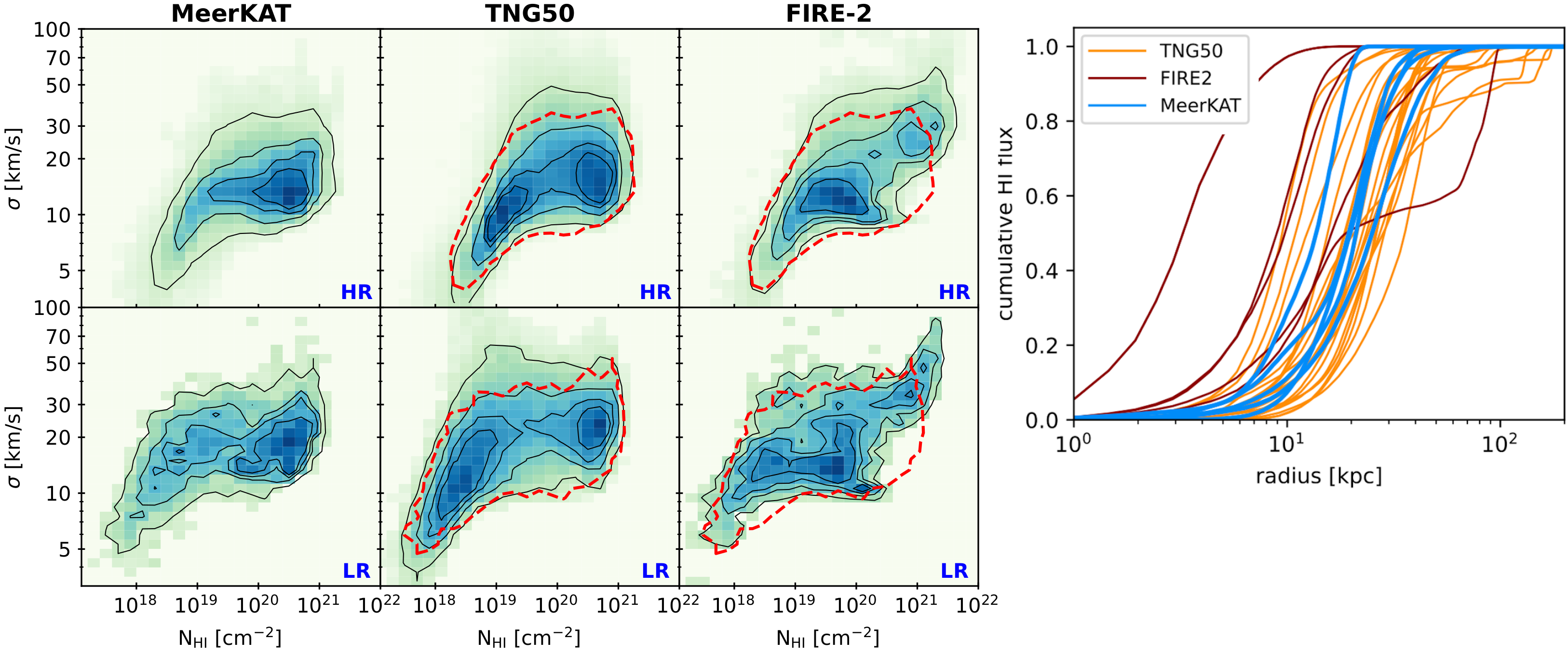}
    \caption{{\it Left panels:} Joint distribution of HI column density ($N_\mathrm{HI}$) and velocity dispersion ($\sigma$). Pairs of values are drawn from corresponding pixels in the $0^\mathrm{th}$ and $2^\mathrm{nd}$ moment maps of high- ($\sim 25^"$; top row -- HR) and low-resolution ($\sim 65^"$; bottom row -- LR) spectral cubes obtained with MeerKAT (left column) for a sample of $5$ $\mathrm L_\star$ galaxies. The centre and right columns show the same measurement for samples of galaxies of similar mass drawn from the TNG50 and FIRE-2 simulation suites, respectively. Simulated galaxies have been `observed' at the same resolution and sensitivity as the corresponding MeerKAT observations. These simulations have much more gas at low column density ($N_\mathrm{HI}<10^{19}\,\mathrm{cm}^{-2}$) and velocity dispersion ($\sigma<15\,\mathrm{km}\,\mathrm{s}^{-1}$) than is seen by MeerKAT, and a tail of high-velocity dispersion gas ($\sigma>30\,\mathrm{km}\,\mathrm{s}^{-1}$) at higher column ($N_\mathrm{HI}>10^{19}\,\mathrm{cm}^{-2}$) that is also not observed. (Reproduced from Fig.~5 in \citealp{Marasco.2025}.) {\it Right panel:} Cumulative HI flux as a function of projected distance for MeerKAT galaxies, compared with selected TNG50 and FIRE-2 galaxies (Reproduced from Fig.~7 in \citealt{Marasco.2025}).}
    \label{fig:sig-nhi}
\end{figure}

Recent work by \citet{Marasco.2025} has highlighted some important differences between cosmological hydrodynamical simulations (here TNG50 and FIRE-2) and galaxies observed during the MHONGOOSE survey in the space of velocity dispersion and HI column density (see Fig.~\ref{fig:sig-nhi}). They highlight two distinct differences. First, there is a `tail' of high velocity dispersion ($\sigma\gtrsim 30\,\mathrm{km}\,\mathrm{s}^{-1}$) gas at higher column densities ($N_\mathrm{HI}\gtrsim 10^{19}\,\mathrm{cm}^{-2}$) in simulated galaxies that is not observed in MHONGOOSE. Second, the simulations seem to predict much more low column density gas ($N_\mathrm{HI}\lesssim 10^{19}\,\mathrm{cm}^{-2}$), mostly residing at lower velocity dispersions, than is observed. The presence of this low-density gas seems to be related to the presence of companion or satellite galaxies -- simulated galaxies that are more isolated are more similar to their observed counterparts. The abundant cold, HI-rich filaments predicted by simulations \citep{Dave.1999,Keres.etal.2005,Ramesh.2023,deBlok.2024,Marasco.2025} do not seem to be observed. They may still be present at lower column densities than predicted, and the SKA could reveal this \citep{deBlok01.2026.SKA}, but for the moment the exact means by which galaxies replenish their atomic gas reservoir remains unclear.

\begin{figure}
    \centering
    \includegraphics[width=0.6\linewidth]{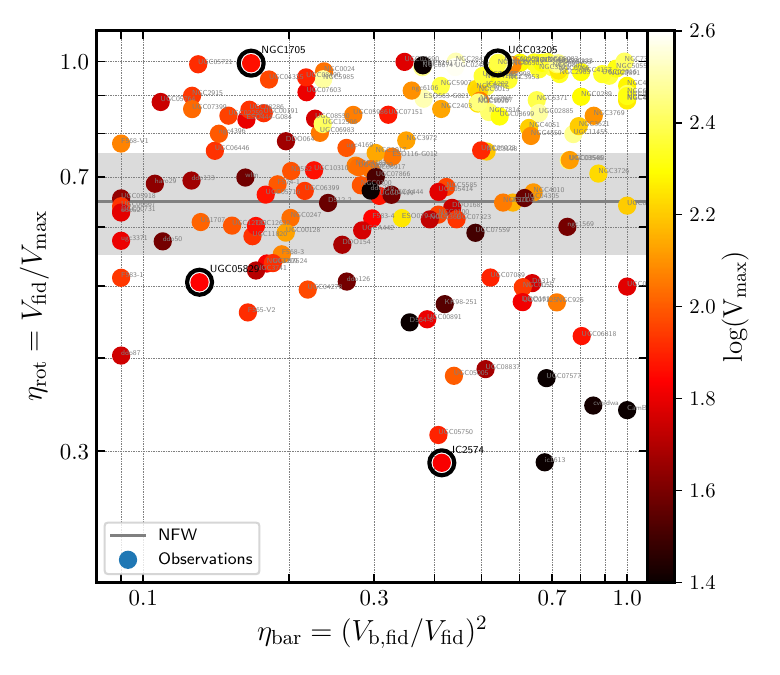}
    \caption{A compilation of observed galaxies collected by \citet{SantosSantos.2020} in the space of $\eta_\mathrm{rot}$ and $\eta_\mathrm{bar}$. The $\eta_\mathrm{rot}$ parameter is the ratio of the rotation curve amplitude at a suitably chosen `inner' radius ($V_\mathrm{fid}$) and its maximum ($V_\mathrm{max}$), and serves as a proxy for the rotation curve shape: a higher (lower) value of $\eta_\mathrm{rot}$ is suggestive of a dark matter cusp (core). The horizontal grey line marks the expectation for an NFW halo model. The $\eta_\mathrm{bar}$ parameter is a measure of the influence of the baryonic (stellar \& gas) components of the galaxy on its kinematics within the same inner aperture. Higher $\eta_\mathrm{bar}$ values correspond to more centrally baryon-dominated galaxies. Points are coloured by $V_\mathrm{max}$. For low mass galaxies ($V_\mathrm{max}\lesssim100\,\mathrm{km}\,\mathrm{s}^{-1}$;black-orange points) an anti-correlation with substantial scatter between $\eta_\mathrm{rot}$ and $\eta_\mathrm{bar}$ emerges. This trend is challenging to explain and provides a stringent test of theories proposed to resolve the cusp-core discrepancy. (Reproduced from Fig.~4 in \citealp{SantosSantos.2020}.)}
    \label{fig:eta-eta}
\end{figure}

Rotation curves extracted from HI observations and galaxy mass models constructed therefrom remain important constraints on the internal structure of galaxies, including through scaling relations such as the baryonic Tully-Fisher relation \citep{McGaugh.2000,ManceraPina.2019}, angular momentum-mass, or Fall, relation \citep{Fall.2013,Posti.2018} and radial acceleration relation \citep{McGaugh.2016,Lelli.2017}. Understanding the diverse shapes of rotation curves, especially those of low-mass galaxies (e.g. Fig.~\ref{fig:eta-eta}), remains a compelling open problem \citep{Oman.2015,SantosSantos.2020}. A generic prediction of dissipationless N-body simulations is that dark haloes should have a dense central cusp with a density slope of approximately $\propto r^{-1}$. Mass models constructed for many galaxies instead suggest constant density central dark matter cores \citep[reviewed by][]{deBlok.2010}. These could be reconciled by (i) redistribution of dark matter in the process of galaxy formation \citep{NEF.1996,Pontzen.2014}; (ii) redistributing dark matter through self-scattering interactions \citep{Spergel.2000,Kaplinghat.2020}; (iii) allowing for a different gravitational force law \citep{Milgrom.1983}; or (iv) appealing to systematic errors in observations or modelling \citep{Oman.2019,Roper.2023}. Which explanation should be favoured remains debated. More generally, reliable inference of the total (dynamical) matter distribution from HI kinematics is essential to many lines of research in galaxy formation and evolution \citep[see][for reviews]{Bullock.2017,Sales.2022}.

Hydrodynamical simulations have a crucial role to play in settling such questions. Forward models of the HI morpho-kinematics of galaxies drawn from simulations \citep[e.g.][]{martini.2024} have provided a powerful means to test physical models against observations \citep{Read.2016,Maccio.2016,Oman.2019,Rey.2024}. Some models become much more challenging to separate once observational effects are accounted for where strong degeneracies are at play \citep{Read.2016,Marasco.2018,Roper.2023}. Simulations have also been used to motivate that the assumption of dynamical equilibrium in gas discs may hold less widely than routinely assumed \citep{Downing.2023,Sands.2024,Dado.2026}.

An enduring observational challenge is that highly spatially and spectrally resolved HI observations are only available for a heterogenously selected sample of galaxies; the SKA can help to alleviate this with larger and crucially more representative samples. Unprecedentedly large samples of marginally resolved HI observations of galaxies are also becoming available now from the WALLABY survey \citep{Murugeshan.2024}, driving the development of modelling tools optimised for this difficult regime and automated to construct large numbers of models (e.g. \citealt{Deg.2025}\footnote{\url{https://github.com/NateDeg/3KIDNAS}}; \citealt{Kamphuis.2015,Haubner.2024}). Hydrodynamical simulations can provide the most powerful test data sets for developing and testing these modelling pipelines, although this approach is so far not in widespread use.

\subsection{HI as a cosmological probe}
%Claudia, Kyle, Connor, Romeel, Gabriella, Lizhi

\begin{figure}
    \centering
    \includegraphics[width=\linewidth]{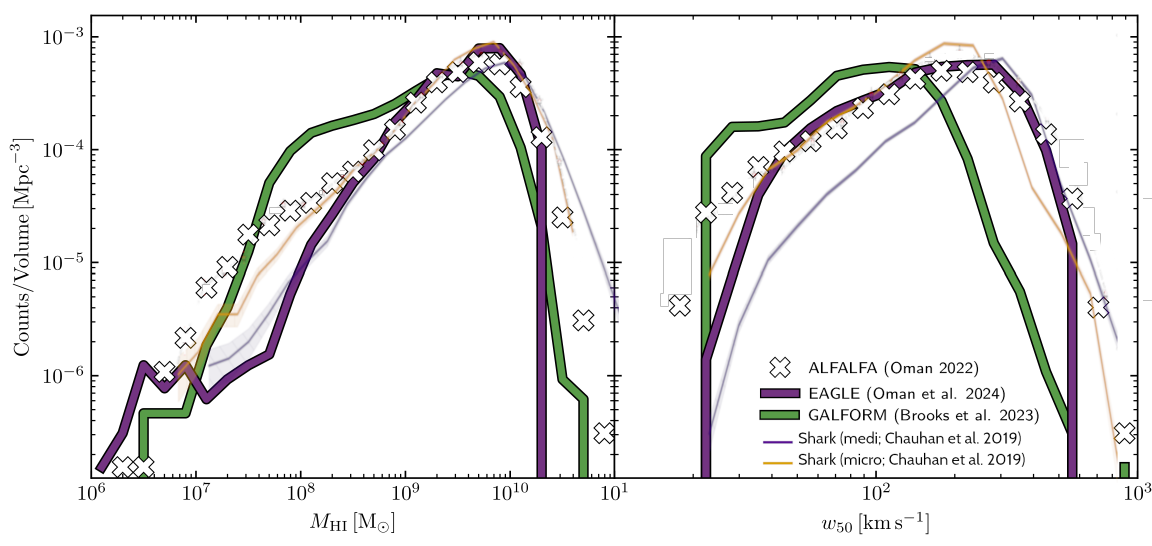}
    \caption{The HI velocity width function is a strong constraint on the halo mass function of gas rich galaxies and therefore a potential test of cosmological models, but faithfully modelling the galaxy distribution in HI mass ($M_\mathrm{HI}$) and line width ($w_{50}$) requires capturing the complexity of cold gas dynamics. A detailed forward model of the spectra of galaxies in the GALFORM semi-analytic model \citep[green line, adapted from fig.~4 in][]{Brooks.2023} greatly overpredicts (underpredicts) the number counts of low-$w_\mathrm{50}$ (high-$w_{50}$) galaxies in the ALFALFA survey \citep[open crosses;][]{Haynes.2018}. However, a similar forward modelling applied to a different semi-analytic model ({\sc Shark}; \citealt{Chauhan.2019}, shown for two $N$-body boxes medi and micro, representing lower and higher resolution, respectively) is in much better agreement with the data, showing how sensitive this is to the underlying baryon physics.     
    Hydrodynamical models like {\sc Eagle} \citep[purple line, reproduced from `EAGLE25' curve in fig.~4 of][]{Oman.2024} capture the non-linear relationship between halo mass and $w_{50}$ -- and especially its scatter -- arguably more accurately \citep[see fig.~7 in][]{Oman.2024}.}
    %Explicitly modelling the cold gas phase, as is done in the {\sc Colibre} model, further improves the agreement (the orange line for {\sc Colibre} is computed analogously to that for {\sc Eagle}, but using the L050m6 fiducial AGN {\sc Colibre} simulation).}
    \label{fig:hiwf}
\end{figure}

\noindent {\it HI velocity function.} The HI velocity width function (HIWF) is the measurement of the number density of HI sources as a function of their HI spectral line width, most often the full width at half maximum ($w_{50}$). Because HI emitting gas extends to large radii in galaxies (compared to the optical component) the HIWF provides a good, but still imperfect, tracer of the total dynamical mass function of gas-rich galaxies that can be used to test predictions for the dark matter halo mass function (HMF) in cosmological models. Indeed, the first measurement of the HIWF \citep[][see also \citealp{Zwaan.2010} for a contemporary measurement]{Zavala.2009} tentatively attributed its shallow low-velocity slope to the suppressed formation of small-scale structure expected in warm dark matter models. This conclusion was debated back and forth as other effects influencing the low-velocity slope were better understood \citep{Papastergis.2011,Trujillo-Gomez.2011,Obreschkow.2013,Sawala.2013,Klypin.2015,Brook.2015,Jones.2015,Brook.2016,Maccio.2016,Ponomareva.2016,Brooks.2017,Schneider.2017,Dutton.2019,Chauhan.2019,Sardone.2024,Oman.2024}. Interest has waned somewhat in recent years but the HIWF remains a potentially powerful constraint on cosmology via the HMF, provided that galaxy formation (especially of dwarfs) is sufficiently well understood \citep{Oman.2022}. Simulations are a powerful tool to encode this coupling between galaxy formation and cosmology. Inferring the HMF from a HIWF measurement is an underconstrained problem \citep{Oman.2024}; a forward modelling approach is a more promising path forward (see \citealp{Obreschkow.2009,Obreschkow.2013,Chauhan.2019,Brooks.2023} for examples based on semi-analytic models, and \citealp{Maccio.2016,Brooks.2017,Dutton.2019,Oman.2024,Sardone.2024} for examples based on hydrodynamical simulations). Semi-analytic galaxy formation models struggle to capture the complexities of gas dynamics that give rise to the HIWF, examples based on the GALFORM and {\sc Shark} models are shown in Fig.~\ref{fig:hiwf}. The two model predict very different distributions, which signals the dependence on the modelling of baryon physics. 
Hydrodynamical simulations fare better in this regard (e.g. the {\sc Eagle} model shown in Fig.~\ref{fig:hiwf}), but until recently still struggled to reproduce the low-velocity end of the HIWF, predominantly by failing to model the ISM's cold gas phase sufficiently accurately. The latest models (e.g. {\sc Colibre}) reproduce the HIWF across the entire measured velocity range to better than a factor of $2$ accuracy (private comm. Kyle Oman), offering the possibility to construct tests relying on subtle signatures such as the influence of a warm dark matter particle. Hydrodynamical simulations cannot simultaneously achieve the high resolution needed for the low-velocity end of the HIWF and the large volumes needed to sample the high-velocity end. Semi-analytic models may have an important role to play in bridging this gap, for instance forward-modelled spectra from hydrodynamical models could be assigned to galaxies from a semi-analytic model to generate full-scale mock surveys.

{\sc HI clustering.} It has become clear over the last decade that cosmological hydrodynamical simulations and semi-analytic models are capable of reproducing very well the HI clustering measurements coming from HIPASS \citep{Meyer.2007} and ALFALFA \citep{Papastergis.2013}; e.g. see \citet{Crain.2017, Zoldan.2017, Chauhan.2020, Fontanot.2025} for examples of the predicted clustering of HI selected galaxies in the {\sc Eagle} cosmological hydrodynamical simulations and the {\sc Shark} and GAEA semi-analytic models, respectively. The key characteristic there is the low bias compared to the clustering of galaxies selected by their stellar mass or luminosity, showing that HI-selected galaxies tend to live in less dense environments. 

However, the predicted evolution of the bias is very different between simulations. For example, \citet{Baugh.2019} using the {\sc GALFORM} semi-analytic model predicted the bias of HI selected galaxies to remain $\lesssim 1$ out to $z\approx 2$, while \citet{Spinelli.2020} using the GAEA semi-analytic model and \citet{Villaescusa-Navarro.2018} using the TNG hydrodynamical simulations predicted the bias to be $>1$ at $z\gtrsim 1$. The difference between these predictions is related to the different evolution they predict for the HI-galaxy-halo mass relation. We discuss this in the next section.

\subsection{Testing baryon physics with HI}

%Feedback effects in the baryon cycle (Ruby, Kyle) -- outflows and accretion
%%% Ruby on feedback/baryon cycle/HI %%%
\noindent {\it HI as a tracer of the baryon cycle.} As described in the introduction, large-volume hydrodynamical simulations are typically calibrated to reproduce the $z=0$ stellar properties of galaxies (e.g. the galaxy stellar mass function, GSMF), leaving their gaseous properties as genuine predictions. Achieving agreement with these stellar observables requires invoking energetic feedback mechanisms from (i) star formation, which flattens the low-mass slope of the GSMF (e.g. \citealt{White.Frenk.1991,Benson.2003}), and (ii) black hole accretion, which produces the characteristic exponential decline at high masses (e.g. \citealt{Croton.2006,Bower.2006}). Owing to the finite mass resolution of baryonic elements — typically between $10^{5}-10^{7} {\rm M}_{\odot}$ — these processes cannot be modelled {\it ab initio}, necessitating a phenomenological sub-grid treatment.

The properties and abundance of atomic gas are extremely sensitive to the manner in which these sub-grid feedback processes are implemented, both within (e.g. \citealt{Crain.2017,Dave.2020}) and external to (e.g. \citealt{Davies.2019,Wright.2020,Crain.2023} galaxies. This is also true of semi-analytic models of galaxy formation (see the comparison in \citealt{Mitchell.2018}). The inherent uncertainty in modelling unresolved feedback processes introduces significant degeneracy in terms of how the stellar mass of galaxies is formed in the context of the broader baryon cycle (e.g. \citealt{Ayromlou.2023,Wright.2024}). \citet{Wright.2024} show, for instance, that stellar-feedback-driven outflows in {\sc Eagle} can reach several times the halo virial radius (hundreds of kpc) in low mass systems ($M_{\rm 200c}\lesssim10^{11.5}{\rm M}_{\odot}$), even entraining mass through the CGM. Comparatively in TNG, while mass loading is higher at the scale of the ISM compared to {\sc Eagle}, recycling typically occurs within the CGM and leads to much higher halo-wise gas fractions than in {\sc Eagle}. 

\begin{figure}
    \centering
    \includegraphics[trim=0mm 4mm 6mm 10mm, clip,width=0.49\linewidth]{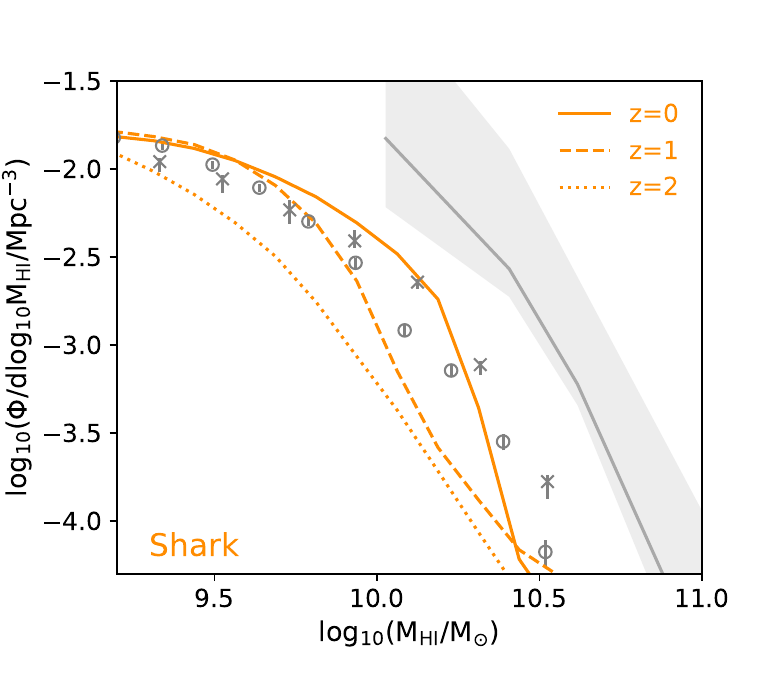}
    \includegraphics[trim=0mm 4mm 6mm 10mm, clip,width=0.49\linewidth]{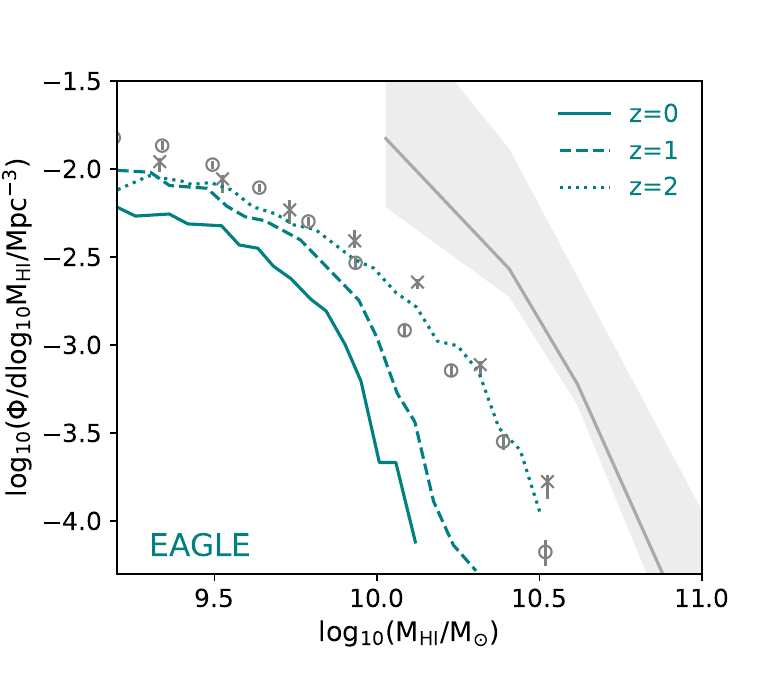}
    \includegraphics[trim=0mm 4mm 6mm 10mm, clip,width=0.49\linewidth]{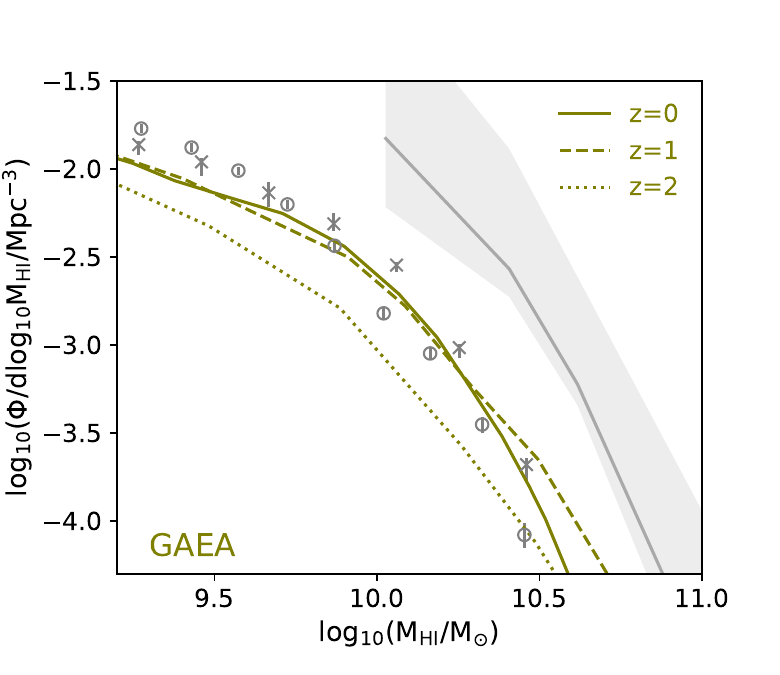}
    \includegraphics[trim=0mm 4mm 6mm 10mm, clip,width=0.49\linewidth]{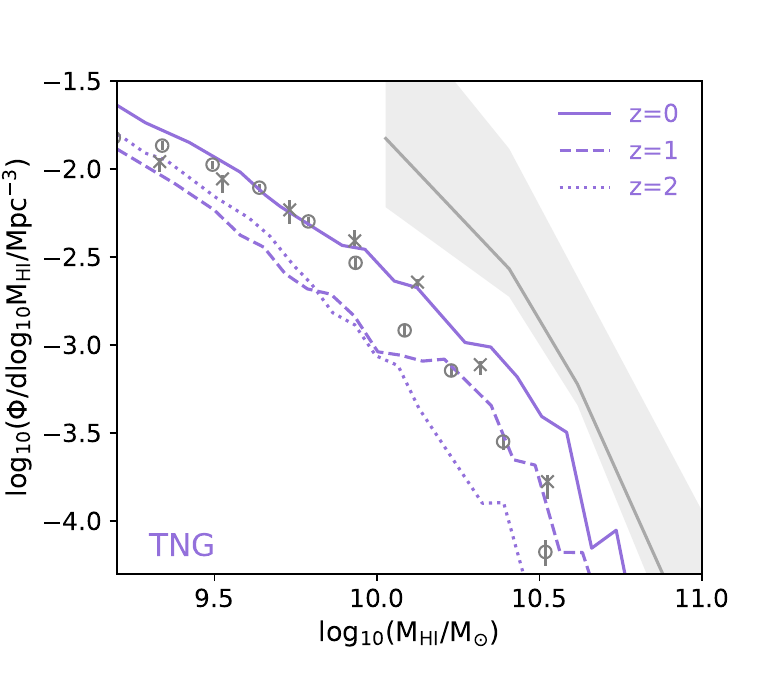}
    \includegraphics[trim=0mm 4mm 6mm 10mm, clip,width=0.49\linewidth]{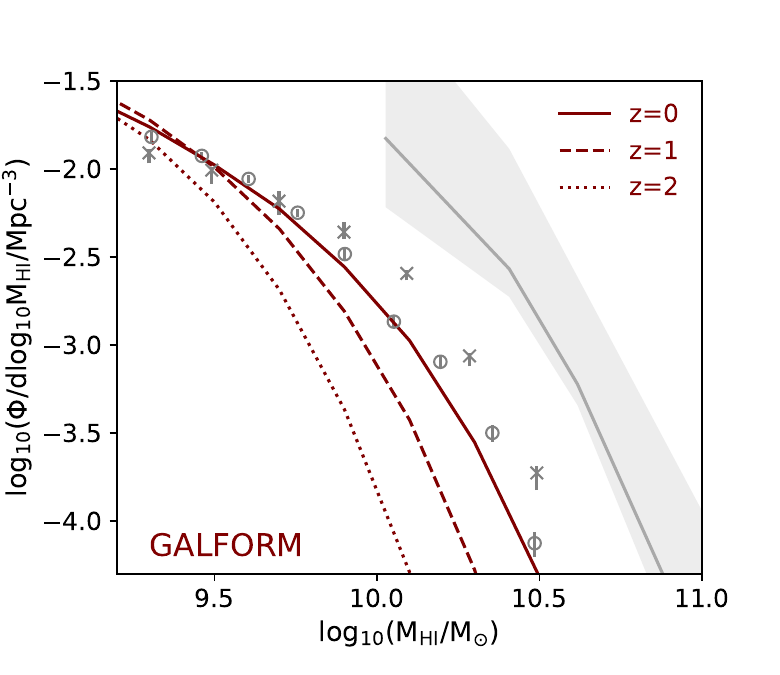}
    \includegraphics[trim=0mm 4mm 6mm 10mm, clip, width=0.49\linewidth]{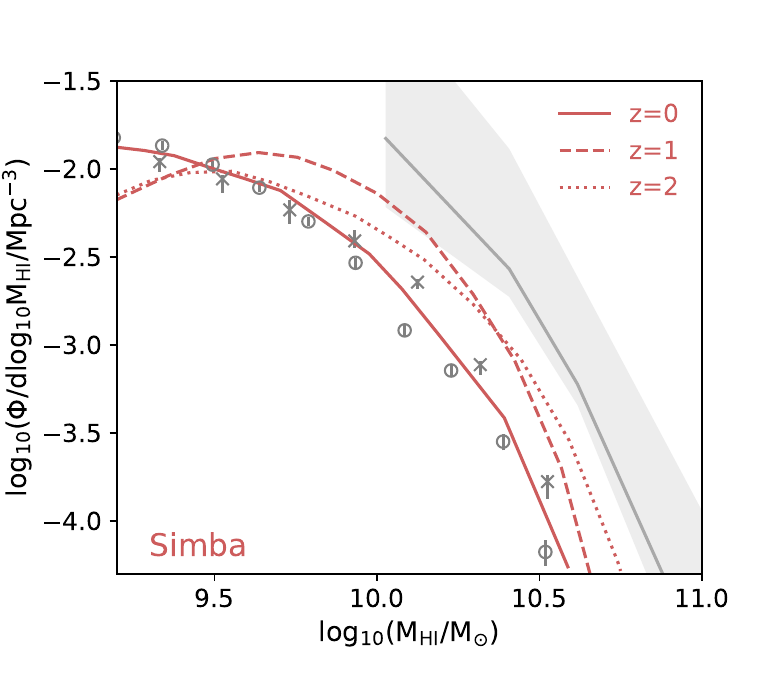}
    \caption{The HI mass function at $z=0$ (solid lines), $z=1$ (dashed lines), and $z=2$ (dotted lines) in $3$ semi-analytic models ({\sc Shark}, GAEA, GALFORM) and $3$ cosmological hydrodynamical simulations ({\sc Eagle}, TNG and Simba), as labelled in each panel. Observations at $z\approx 0$ from \citet{Zwaan05} and \citet{Jones.2018} are shown as grey circles and crosses, respectively; while the 
    inferred HI mass function at $z\approx 1$ from convolving stacked spectra with measured optical luminosity functions  at  $z\approx 1$ \citep{Chowdhury.2024} are shown as a grey band. {\sc Eagle} and Simba predict the number densities of galaxies to increase with increasing redshift, while the other simulations predict the opposite trend. None of the simulations appear to reproduce the level of increase in number density inferred from the $z\approx 1$ observations. }
    \label{fig:HImfz0}
\end{figure}

\begin{figure}
    \centering
    \includegraphics[width=0.9\linewidth]{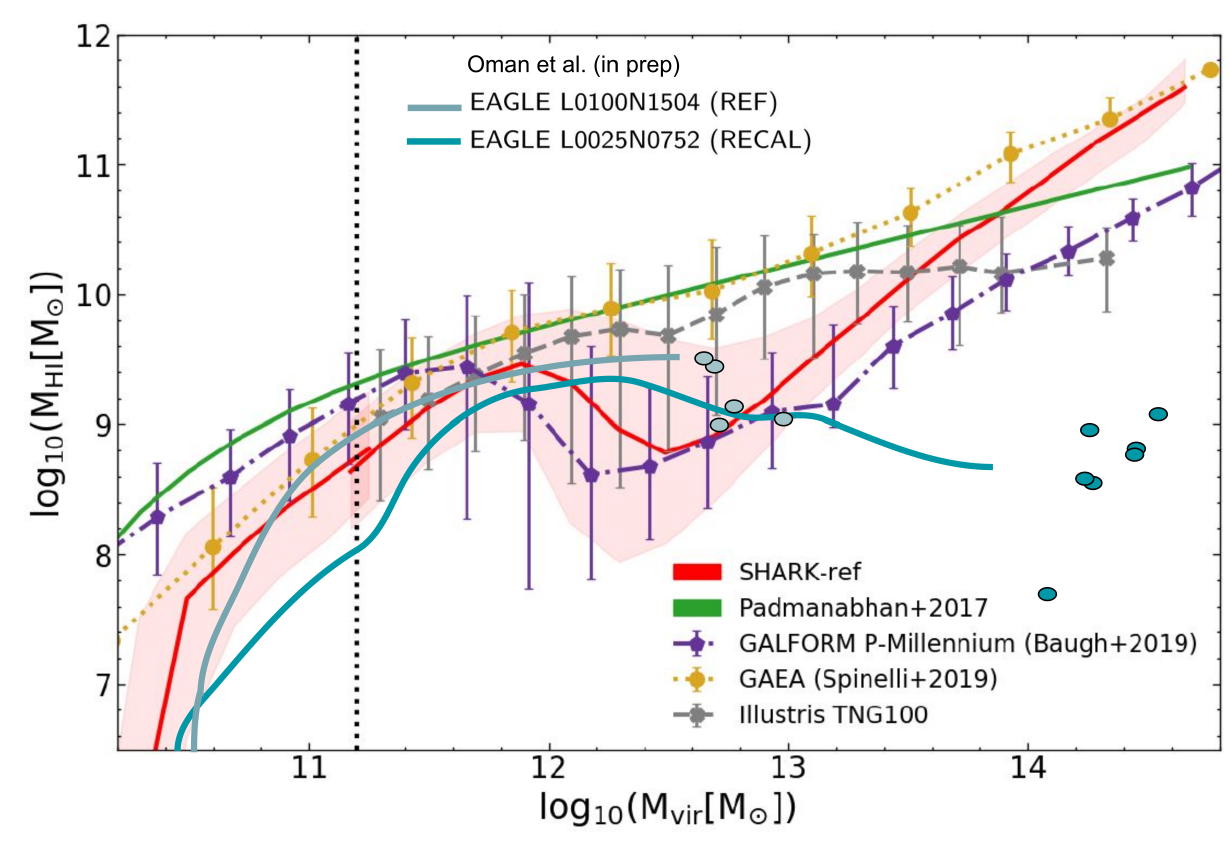}
    \caption{The HI mass vs halo mass relation at $z=0$ in $3$ cosmological hydrodynamical simulations (TNG and two {\sc Eagle} realisations) and $3$ semi-analytic models of galaxy formation ({\sc Shark}, {\sc GAEA} and {\sc GALFORM}). Also included is the empirical results of \citet{Padmanabhan17} for reference. For {\sc Eagle} we also show individual massive halos with symbols in the regime where medians cannot be calculated due to the low number of objects. Figure has been modified from \citet{Chauhan20}.}
    \label{fig:HIHMz0}
\end{figure}

Applying a post-processing-based decomposition of the neutral gas into its constituent species allows one to see how the difference in total gas content translates to differences in halo HI content in these models, as demonstrated in the $M_{\rm 200c}-M_{\rm HI}$ relation (discussed below and in Fig. \ref{fig:HIHMz0}). It is clear that TNG predicts $M_{\rm HI}$ several times that in {\sc Eagle} at fixed halo mass (below $M_{\rm 200c}=10^{12}\, {\rm M}_{\odot}$), simply due to the presence (or lack thereof) of recycling HI, while both still producing a very similar local HIMF (as per Fig.~\ref{fig:HImfz0}). While the predicted relation is sensitive to both the post-processing decomposition method (e.g. \citealt{Blitz.2006,Gnedin.2011,Krumholz.2013}, see discussion in \citealt{Lagos.2015, Bahe.2016, Crain.2017} as well as the halo mass measure (e.g. \citealt{Dev.2023}), the HI content of haloes has the potential to serve as a powerful constraint for the sub-grid feedback prescriptions employed in simulations. This is particularly pertinent for the next-generation of cosmological simulations with a cold ISM and on-the-fly decomposition of gas into constituent species (\citealt{Feldmann.2023,Schaye.2025}). 

%\begin{figure}
%    \centering
%    \includegraphics[width=\linewidth]{Figs/%Marasco25_adaptedfig6fig7.png}
%    \caption{Taken from Figures 6 and 7 in \citet{Marasco.2025}. Left panel: the probability density distribution of inclination-corrected $N_{\rm HI}$ in MeerKAT galaxies (blue), compared with mock-observed TNG50 (orange) and FIRE-2 (dark red curves) galaxies. Right panel: Cumulative HI flux as a function of projected distance for MeerKAT galaxies, compared with selected TNG50 and FIRE-2 galaxies.}
    %\label{fig:nhi_pdf_cdf}
%\end{figure}

\begin{figure}
    \centering
    \includegraphics[width=0.99\linewidth]{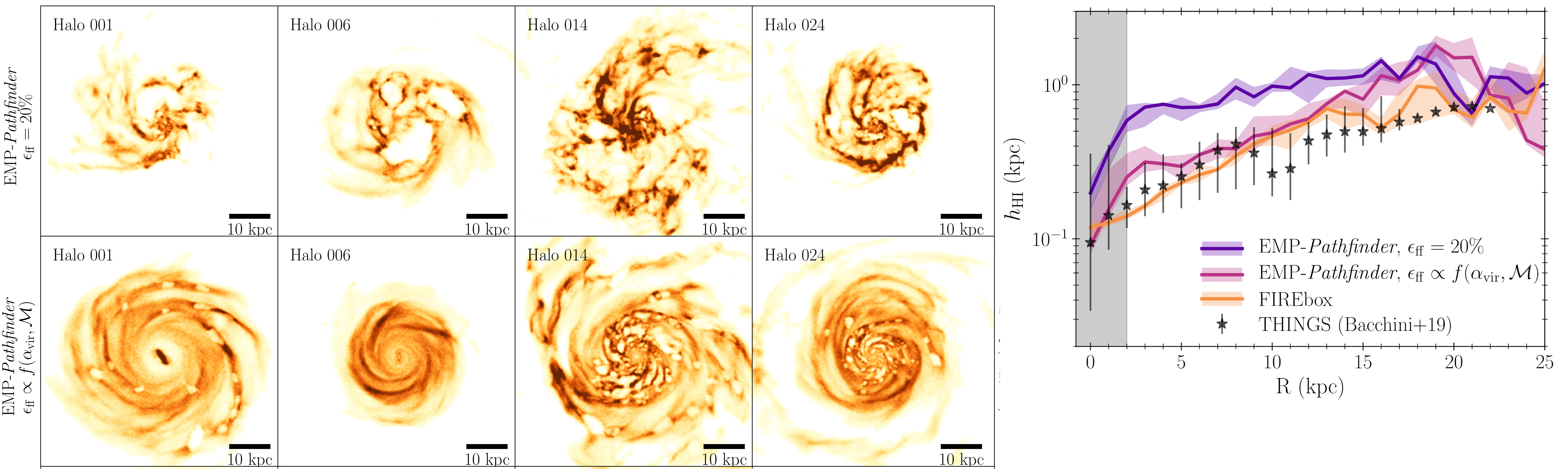}    
    \caption{Taken from Figures 4 and 7 in \citet{Gensior.2024}. The maps on the left show face-on projection of the HI surface density for four example galaxies in the EMP-{\it Pathfinder} simulations employing a constant star formation efficiency (top) and an efficiency that depends on the turbulence of the ISM (bottom). The plot on the right shows radial profiles of the HI disc scale heights of the simulated galaxies, showing that the EMP-{\it Pathfinder} simulations employing a constant star formation efficiency produces HI discs that are substantially thicker than those observed in the nearby universe (black symbols with errorbars).}
    \label{fig:HIscale}
\end{figure}

Detailed studies of the diffuse, low column density HI ($N_{\rm HI}\lesssim10^{18}\,{\rm cm}^{-2}$) surrounding individual galaxies are also an important, and varying, prediction of hydrodynamical simulations. Such predictions have been compared with data from the MHONGHOOSE survey \citep{deBlok.2024}, which provides deep, spatially resolved HI observations of nearby ($D \lesssim 20\, {\rm Mpc}$) galaxies with the MeerKAT telescope. In addition to the excess HI at low column densities discussed in Section~\ref{sec:kinematics}, 
%
%low HI column The 
%MHONGOOSE vs TNG50 and FIRE-2 comparison of 
%\citet{Marasco.2025} (discussed in Section \ref{sec:kinematics} above), revealed that 
%compare MHONGOOSE data with mock HI observations of galaxies in the TNG50 and FIRE-2 simulations, with a focus on the HI at the disk-CGM interface. The left panel of Fig. \ref{fig:nhi_pdf_cdf}, reproduced from Fig. 6 of \citet{Marasco.2025}, illustrates that galaxies in TNG50 and FIRE-2 exhibit an excess of low-column density HI ($\lesssim 10^{20} {\rm cm}^{-2}$) relative to the observations. The right panel of Fig. \ref{fig:nhi_pdf_cdf} (reproduced from Fig. 7 in \citealt{Marasco.2025}) indicates where this gas resides, displaying the cumulative HI flux from the selected galaxies as a function of projected distance. In 
\citet{Marasco.2025} also showed TNG50 produced a significantly higher fraction of HI distributed at large distances (50-150\,kpc) from the galactic centre — up to $14\%$ beyond 70~kpc, while the maximum in the observed galaxies is always $<1\%$ (see right panel in Fig.~\ref{fig:sig-nhi}).

This tension with observations is in line with the picture of efficient intra-halo recycling and high CGM $f_{\rm gas}$ present in TNG found in \citet{Wright.2024}. The presence of circulating gas in the inner CGM of these simulated galaxies is also offered as an explanation for the relative broadness of the mock-observed HI line profiles compared to MeerKAT observations, with the authors arguing that gas cycling in the inner CGM may be too ``vigorous'' in the simulations relative to observations. With SKA-mid slated to be able to probe down to column densities $\approx 0.5{\rm dex}$ lower than MeerKAT (at 100h observation time), careful comparisons between mock-observed galaxies employing different stellar and AGN feedback prescriptions is a very promising avenue to constrain their uncertain sub-grid feedback models in the SKA-era. 

While we discuss the influence of outflows, recycling and the baryon cycle on the static halo HI content above, direct observations of cold outflows in nearby galaxies are becoming more frequent, with indications that HI dominates the mass loading at the disk-CGM interface (e.g. \citealt{Mazzilli.Ciraulo.2025}). Confronting forward-modelled predictions for multi-phase outflows from the next generation of cosmological simulations with an increasing number of detailed, multi-wavelength observational studies of outflows in individual galaxies has the potential to form a powerful test for the accuracy sub-grid physics models.

High resolution simulations have been used to demonstrate that the modelling of star formation and stellar feedback impacts the distribution of HI. \citet{Gensior.2024} showed that some models predict HI disks that are too thick compared with observations, and highlighted that the central HI disc morphology is very sensitive to the interaction between the adopted stellar feedback and the star formation models. As shown in Fig.~\ref{fig:HIscale}, whether star formation is modelled with a constant efficiency or tied to the self-gravity of the gas affects the location and density structure of the gas where feedback events occur and thus its effect on the HI disc structure (cf. the two different EMP-{\it Pathfinder} samples). Similarly, different stellar feedback treatments (cf. EMP-{\it Pathfinder} with FIREbox that includes various early stellar feedback channels) can lead to systematic differences in e.g. the thickness of HI disks, due to early stellar feedback pre-processing the gas reservoir in which supernovae explode. Fig.~\ref{fig:HIscale} highlights how quantitatively comparing the resolved HI distribution of a statistical sample of simulated galaxies with observations, can be used to constrain subgrid physics models. 

%%%%%%%%%%%%%%%%%%%%%%%%%%%%%%%%%%%%%%%%%%%%%

%Feedback effects in the HI-halo mass relation (Claudia, Chris, Gabriella, Lizhi) -- AGN feedback, others
\begin{figure}
    \centering
    \includegraphics[width=\linewidth]{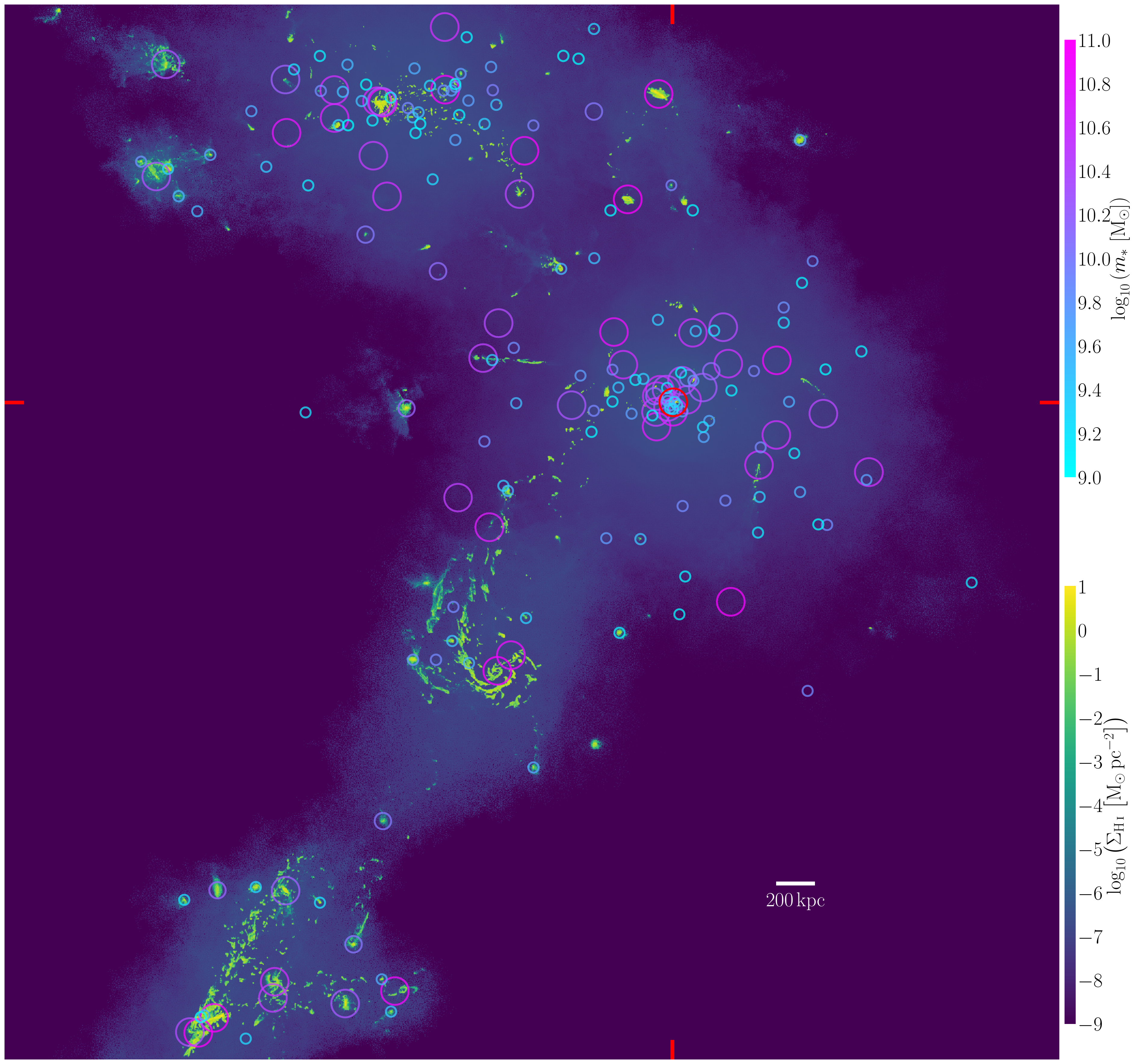}
    \caption{Figure 1 in \citet{Stevens.2019}, showing the projected HI column density of a galaxy cluster of mass $1.7\times 10^{14}\,\rm M_{\odot}$ in TNG at $z=0$.  Circles indicate the equivalent physical beam size used to measure HI masses considering the xGASS survey \citep{Catinella.2018}. Circles are coloured by the stellar mass of the galaxy
 they are centred on. The central galaxy of the cluster is indicated by red circle.}
    \label{fig:group3}
\end{figure}

\noindent {\it Environmental effects traced by the HI morphology and content of galaxies.} 
%(Danail, Romeel)
%
%[text from Lizhi]
Galaxies in cluster environment are observed  to be HI deficient with respect to isolated galaxies of the same optical size (e.g. \citealt{Cortese.2011,Li.2020}). These observations show that the HI deficiency increases towards the cluster centre. Interestingly, along with the reduced HI content, the molecular hydrogen reservoir also appears to be (mildly) affected \citep{Boselli.2014}. \citet{Stevens.2017} and \citealt{Xie.2018} using the semi-analytic models Dark Sage and GAEA, respectively, showed that direct stripping of cold gas was needed to reproduce the observed trends towards clusters as well as statistical trends of satellite galaxies becoming more gas poor at higher halo masses. \cite{Chen.2024} compared the semi-analytic models L-Galaxies and GAEA, and the hydrodynamical simulations {\sc Eagle} and TNG, with WSRT data measuring the HI content and SFR of galaxies across projected cluster centric radii. All models predicted the same qualitative trend of the HI mass of galaxies decreasing towards a halo's center, as it is observed. However, satellite galaxies from hydro-dynamic simulations are generally found to be too gas poor compared both with semi-analytic models and observations across all projected radii. This indicates that the environmental stripping of cold gas in {\sc Eagle} and TNG is much too efficient. Interestingly, it appears that the difference in HI stripping efficiency is not caused by different ram pressure stripping strengths in hydro-simulation and semi-analytic models. In fact, \cite{Xie.2025} investigated the predicted ram pressure strength in the phase space of cluster halos and found that the values were broadly consistent between the TNG and GAEA models, even though the semi-analytic model uses simplified algorithms to compute ram pressure. 
A critical aspect is likely related to the limited resolution in hydrodynamical simulations, which becomes especially serious when analysing how galaxies lose gas as they orbit groups and clusters \citep{Wright.2022}.  

An important caveat to consider when comparing simulations with HI observations of galaxies in dense fields are the typically large beams which can lead to larger HI reservoirs being associated with a single galaxy. \citet{Stevens.2019} using the TNG hydrodynamical simulations demonstrated the impact that can have on the recovered HI masses of galaxies in clusters. This is shown in Fig.~\ref{fig:group3} for a cluster in TNG at $z=0$. The underlying HI maps can be very complex, which sometimes makes it hard to determine which gas to associate with a galaxy. In any case, this serves as a demonstration for the importance of forward modelling when compared with observations. 

\noindent {\it The HI mass function.}
 Fig.~\ref{fig:HImfz0} shows the predicted evolution  of the HI mass function from $z=0$ to $z=2$ in $6$ different simulations ($3$ semi-analytic models and $3$ cosmological hydrodynamical simulations). 
At $z=0$, all the simulations are in reasonable agreement with observations,  
%The top panel of Fig.~\ref{fig:HImfz0} shows the $z=0$ HI mass function predicted by 6 simulations. 
%
%They all agree relatively well with observations at $z=0$, 
except for the {\sc Eagle} simulations. However, for {\sc Eagle} this is primarily driven by resolution, as it was demonstrated in \citet{Crain.2017} and \citet{Dave.2020}. 

This level of agreement among simulations and with the observations at $z=0$ is not necessarily surprising given the simulations are fitted to reproduce $z=0$ observables related to the stellar mass and star formation rates of galaxies, which are indirectly related with the HI content of galaxies. Additionally, in the case of {\sc GAEA} the $z=0$ HI mass function is explicitly used as one of the constraints to fit the free parameters of the model. Note that the current $z=0$ observations are quite precise (which can be seen by the small errorbars), which would imply a significant tension with some of the models. However, the observations themselves are in significant tension with each other, potentially implying important systematic effects. One of the challenges of current $z=0$ observations is the relatively small volumes probed due to the shallowness of the surveys. This will dramatically change with the SKA, which will probe large cosmological volumes at much greater depth than existing HI surveys, providing unparallel constraints of the atomic hydrogen content of the local Universe.  

The most striking differences between the simulations become evident as one compares what they predict for the evolving HI mass function. {\sc Eagle} and Simba predict a significant increase (of $\approx 0.3-0.5$~dex) in the number density of galaxies across almost the whole HI mass range probed between $z=0$ and $z=1$. This contrasts with what is predicted by {\sc Shark}, GALFORM and TNG, that show a decrease in the number density of galaxies, particularly at HI masses $\gtrsim 10^{10}\,\rm M_{\odot}$, from $z=0$ to $z=1$. GAEA on the other hands predicts almost no evolution between $z=0$ and $z=1$. 
%The middle panel of Fig.~\ref{fig:HImfz0} shows the the predicted HI mass function now at $z=1$. Most simulations predict either no evolution or an overall decrease in the number density of HI massive galaxies from $z=0$ to $z=1$, while Simba predicts the opposite trend, with the abundance of HI massive galaxies increasing from $z=0$ to $z=1$. 
Recently, tentative measurements of the HI mass function at $z=1$ presented in \citet{Chowdhury.2024} imply an increase in the number density of galaxies with HI masses $\gtrsim 10^{10}\,\rm M_{\odot}$ of one order of magnitude relative to $z=0$. If confirmed, this would be in significant tension with most predictions, with Simba being the only simulation to come close to that strong evolution.
%prefer higher number densities of HI massive galaxies compared with the observed $z=0$ HI mass function. potentially leaning towards the predictions of the Simba simulations.

Going from $z=1$ to $z=2$, some simulations predict a continuous decrease in the number densities of HI massive galaxies, such as is the case in {\sc Shark}, GAEA, GALFORM, Simba. TNG on the other hand predicts little evolution except for the most massive galaxies, while {\sc Eagle} predicts a large increase in the abundance of HI rich galaxies, of $\gtrsim 0.5$~dex. This demonstrate that the simulations are predicting very different evolutions of the HI gas content of galaxies across cosmic time, which will be probed to an unprecedented level by the SKA. It is clear that any constraints on this space would have the ability to rule out several models. 

The different predictions stem again from the different underlying models for the ISM, star formation and feedback, all of which impact the baryon cycle and the instantaneous gas content of galaxies. 
%, but so far very few to no observational constraints are available.

\noindent {\it The HI content of halos.} Although the simulations predict relatively similar HI mass functions at $z=0$, that is not the case for the predicted HI-halo mass relation at $z=0$.
This is shown in Fig.~\ref{fig:HIHMz0}, with a compilation of predicted HI-halo mass relations from different simulations. Here, the y-axis is the total HI content within a halo. The simulations predict disparate relations. In particular, there are some simulations that predict a clear break in the relation at around $10^{12}\,\rm M_{\odot}$ (e.g. {\sc Eagle}, {\sc Shark} and GALFORM), while other simulations predict a monotonic relation between HI and halo mass (e.g. TNG and GAEA). The behaviour of simulations around this scale is very sensitive to how AGN feedback is modelled in simulations, so it is a especially interesting regime to test with observations.

\citet{Chauhan.2020} show that the scatter around the break point of the HI-halo mass relation is strongly correlated with the black hole-stellar mass relation in {\sc Shark}, with overly-massive black holes associated with lower HI content at fixed halo mass. This strengthens the link between this potential break of the HI-halo mass relation with AGN feedback. 
Oman et al. (in preparation) analyse the the impact of the modelling of AGN feedback on the break of the HI-halo mass relation in the {\sc Colibre} simulations. They compare a thermal AGN feeedback model, in which black holes dump thermal energy to the nearest gas particles, heating them to very high temperatures in an isotropic way; and a Hybrid AGN feedback model, which corresponds to two modes of AGN feedback, a thermal and a kinetic-jet mode. These two models are described in detail in \citet{Schaye.2025,Husko.2025}. Oman et al. find that the depression in the HI mass of halos kicks in at lower halo masses in the hybrid model, $\approx 10^{11.7}\,\rm M_{\odot}$ and leads to a deeper depression in the HI gas content of halos, than the thermal AGN feedback model. This exemplifies the sensitivity of the HI-halo mass scaling relation to feedback mechanisms. 

Another important break of the HI-halo mass relation is predicted to happen at lower halo masses, $\approx 10^{10.5}\,\rm M_{\odot}$, as shown by \citet{Villaescusa-Navarro.2018} using TNG, which is associated with halos being strongly affected by stellar feedback and the UV background. 

Although observations of this relation at $z=0$ have been presented in the literature, there are large systematic effects that prevent them from distinguishing between these different predicted scenarios. These systematic effects are primarily driven by the difficulty in measuring halo masses of low mass groups as shown by \citet{Chauhan.2021}. The uncertainties around measurements of halo mass, particularly at low galaxy occupancy, are so large that they tend to wash out the dip in the HI content of halos around the $10^{12}\,\rm M_{\odot}$ halo mass scale to appear like a monotonically increasing relation. \citet{Chauhan.2021} demonstrated that with upcoming deep, highly complete spectroscopy surveys of galaxies (e.g. \citealt{WAVES,4HS}), these uncertainties will decrease significantly by sampling more galaxies per halo mass at the critical $10^{12}\,\rm M_{\odot}$ scale. This shows that there are strong commensalities between the SKA and optical/near-infrared surveys that will need to be strategically exploited to reach the best understanding of galaxy formation and evolution (see \citealt{Duncan01.2026.SKA} for a discussion about potential strategies). 

%%%%%%%%%%%%%%%%%

\section{Recent progress and challenges in modelling radio continuum}\label{sec:contprogress}

\subsection{Modelling the contribution from star formation}
%Claudia, Danail, Filip

The framework for modelling the radio continuum emission associated with star formation has consistently moved from more empirical postprocessing of simulations to more physically motivated models. 
\citet{Thomas.2021} used the Simba cosmological hydrodynamical simulations together with a simple scaling between the $1.4$~GHz continuum and the star formation rate, to predict the radio luminosity functions (RLFs) and showed that the simulation was able to reproduce the RLF associated with star-forming galaxies in the local Universe. They predicted that the vast majority of $1.4$~GHz emitting sources with luminosities $\lesssim 10^{23}\,\rm W/Hz$ are star-forming galaxies, while AGN only dominate at higher luminosities. 

\citet{Hansen.2024} introduced a physically-motivated star formation radio continuum module to the {\sc Shark} semi-analytic model, which included three emitting components: the  thermal free-free emission associated with free electrons emerging from the ionized-gas around star forming regions; synchrotron emission associated with core-collapse supernovae rate; and a small supernovae remnant term. The model was capable of predicting the emission across a wide radio frequency range, and was tested from $150$~MHz to $1.4$~GHz, demonstrating broad consistency with measured RLFs (see top panels in  Fig.~\ref{fig:RLF_Shark}) and observed scaling relations between the radio continuum and far-infrared emission. \citet{Hansen.2024} shows that that the observed evolution in the IR–radio correlation can largely be attributed to an artefact of radio-quiet AGN contamination, while the underlying relation is not evolving.
 Although the modules introduced by \citet{Hansen.2024} are more physical than the scaling relations employed in \citet{Thomas.2021}, they still make some significant simplifications, such as neglecting the dependence of the emissivity on the magnetic field due to its weak scaling and neglecting the time delay associated with supernovae exploding. There is thus a lot of room for improvement in the coming years. 

Despite this valuable progress, it is fair to say that the simulations have moved much more slowly in improving the modelling of radio continuum associated with star formation compared to the effort put in improving the models of the atomic and molecular hydrogen content of galaxies. Some of this is related to the more direct impact the latter has on the baryon cycle and thus, improving the ISM modelling appears more imperative than improving the radio continuum emission of galaxies. This is thus an area in which the community needs to focus in preparation for the SKA. 

\subsection{Modelling the contribution from AGN}
%\textbf{Nicole, Chris, Claudia, Romeel, Cedric, Filip}

The theoretical framework for radio continuum emission from AGN has evolved significantly over the past decade, driven by advances in both empirical modelling and physically motivated simulations. 

Empirical models use scaling relations to assign radio flux densities to galaxies based on star formation rates to recover the statistics of radio AGN in the observed galaxy population. The recent work of \citet{Gao.2025} showed that such an approach can reproduce observed RLF and number counts across multiple bands. They predict that $\approx 50$\% of high redshift radio sources will be optically invisible between $z$=4 and 6, and future SKA radio continuum surveys will be crucial if we are to detect them and better understand early galaxy evolution. The Tiered Radio Extragalactic Continuum Simulation \citep[T-RECS;][]{Bonaldi.2019,Bonaldi.2023} provides  predictions for AGN and star-forming galaxies across a wide frequency range (150 MHz to 20 GHz), as well as HI emission and continuum–HI cross-correlation. This model predicts that the star formation rate–radio luminosity relation depends primarily on stellar mass rather than redshift, consistent with recent observational constraints.

The RAiSE framework \citep[][]{Turner.2018a,Turner.2018b,Turner.2023} develops an analytical model for AGN jet evolution, capturing the full lifecycle from initial jet expansion to remnant phases. It predicts that young sources ($\lesssim$10 Myr) exhibit longer and brighter lobes than previously expected, due to a jet-dominated expansion phase. This has implications for interpreting radio galaxy morphologies and energetics. The PRAiSE extension \citep[][]{Yates-Jones.2022} applies spatially resolved radiative losses to synthetic radio emission, revealing strong redshift dependence due to inverse-Compton losses and suggesting spectral index asymmetry as a tracer of environmental density.

Environmental studies using GALFORM 
\citep[][]{Izuuierdo-Villalba.2018}
show that radio galaxies reside in more massive halos than radio-quiet counterparts, with AGN feedback suppressing stellar mass growth and shaping the infrared luminosity function. These findings suggest that AGN feedback imprints observable signatures on galaxy environments, particularly at high redshift.

\begin{figure}
    \centering
    \includegraphics[width=1\linewidth]{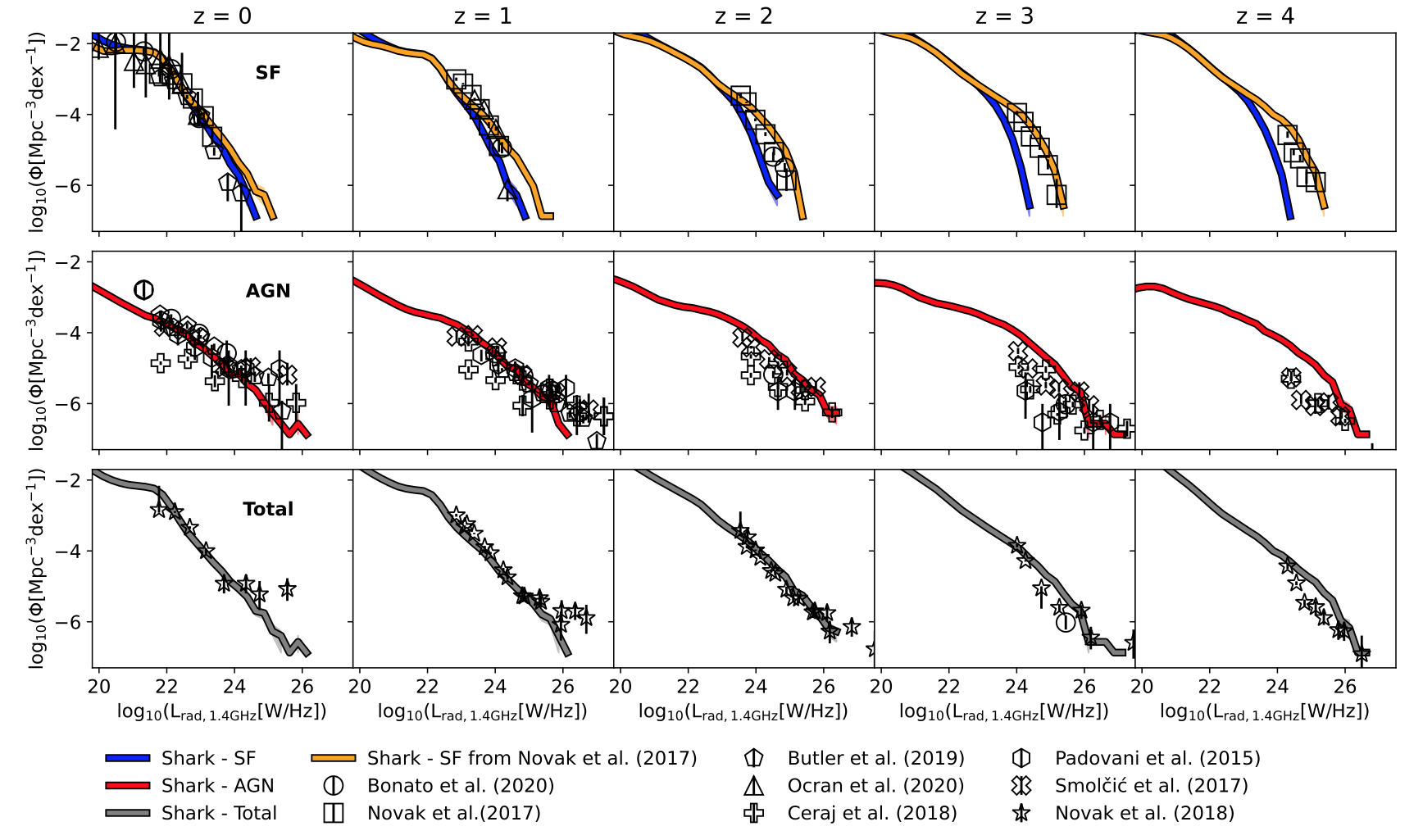}
    \caption{The Radio Luminosity Function (RLF) at 1.4 GHz between $z$=0 and 4 in the {\sc Shark} semi-analytic model \citep{Lagos.2018,Lagos.2024}. The top panels show the contribution from star formation; the middle panels from AGNs; and the lower panels show the combined contributions (i.e. star formation and AGNs). Comparisons are made with observational data of galaxies classified as star forming galaxies in the top panels, AGNs in the middle panels, and the total RLF in the bottom panels. Taken from Figure 3 in \citet{Hansen.2024}.}
    \label{fig:RLF_Shark}
\end{figure}

More recently, \citet{Hansen.2024} included a model for the core radio emission of jets in the {\sc Shark} semi-analytic model that considered all the properties available for black holes in that model (mass, spin, accretion rate and accretion disk geometry). The middle panels in Fig.~\ref{fig:RLF_Shark} show what the model predicts for the $1.4$~GHz RLF associated with AGN from $z=0$ to $z=4$, demonstrating broad agreement with observations. Interestingly, \citet{Hansen.2024} showed that radio quiet AGN can produce enough radio continuum as to be confused with emission coming from star formation. This is because the emission is not enough as to significantly deviate the host galaxy from the far-infrared-radio correlation, generally employed to separate star-forming from AGN galaxies. This is demonstrated by the yellow lines in the top panels of Fig.~\ref{fig:RLF_Shark}, which show the derived RLF if one was to apply the same observational selection of star-forming galaxies. The AGN contamination (the difference between the yellow and blue lines in the top panels) primarily impacts the bright end of the RLF. This again helps to stress the importance of forward modelling when comparing with observations. 

\citet{Thomas.2025} using a model for the radio continuum associated with AGN and applied to the Simba hydrodynamical simulations predict that there exist a population of low Eddington ratio high-excitation AGN sources, which defy the common assumption of low Eddington ratio AGN being associated with low-excitation AGN. Hints of the existence of that population are already present in the MIGHTEE (MeerKAT International GHz Tiered Extragalactic Exploration) radio galaxy population as shown in \citet{Thomas.2025}. With the depths expected for SKA surveys \citep{Hardcastle01.2026.SKA,Duncan01.2026.SKA}, these rare populations should become much better characterised, serving 
%
%They argue that such sources should become blatantly clear at the depths observed with the SKA, and should serve 
to probe the modelling of jets in cosmological hydrodynamical simulations. 

\subsection{Modelling the contribution from shocks associated with halo accretion and mergers}
%\textbf{Filip Husko, Chris Power}
\begin{figure}
    \centering
    \includegraphics[width=1\linewidth]{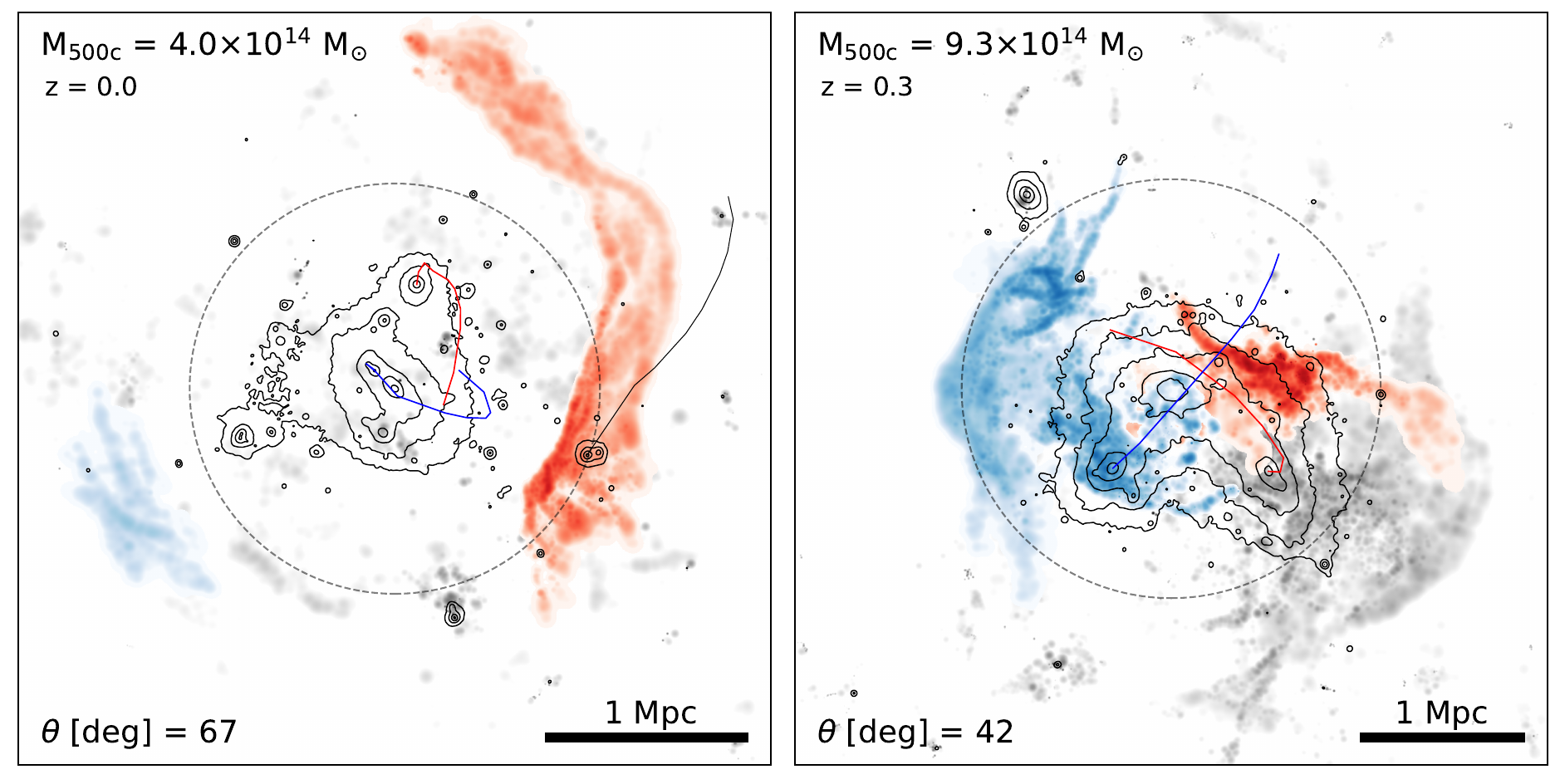}
    \caption{Figure 6 in \citet{Lee.2025} showing radio relics in TNG-Cluster and TNG300 and the complex morphologies they can have. The colormap shows the radio surface brightness, while the primary and the secondary radio relics are shown in red and blue colormaps, respectively. The contours show the dark matter surface density, and the dotted line shows the R500c of the cluster. The red and blue lines show the orbit of the most massive and second most massive sub-cluster, respectively, over the last $1.5$~Gyr. In the left panel example, there is a halo getting close to the merging pair (shown by the black line) which boosts the local surface brightness of the primary radio relics.}
    \label{fig:relics}
\end{figure}
A basic theoretical picture of how gas accretes onto galaxies from the inter-galactic medium (IGM) via their circumgalactic media (CGM) has been established for several decades. Gas is shock heated to the virial temperature, $T_\text{vir}$ of the galaxy's host dark matter halo upon accretion, and after a cooling time it will flow to the halo centre and onto the galaxy \citep{Rees.Ostriker.1977MNRAS,Silk.1977,White.Rees.1978MNRAS,White.Frenk.1991}. While higher mass halos can form long-lived virial shocks, lower mass ones cannot, and this allows for rapid transport via the CGM onto the galaxy on a free-fall time \citep{Birnboim.Dekel.2003}. These modes of hot and cold accretion have subsequently been characterised in hydrodynamical simulations, and this theoretical picture has been refined. The preferential accretion of cold gas along filaments can produce infalling gas streams with high densities and short cooling times \citep[e.g.][]{Keres.etal.2005,VanDeVoort.etal.2012}, which can penetrate hot atmospheres of higher mass halos.

\citet{Theuns.2021} highlighted the importance of cosmological gas accretion for neutral hydrogen in the CGM, demonstrating that Lyman Limit Systems (LLSs) and Damped Lyman-$\alpha$ (DLAs) systems at $z\approx 2$ are not consistent with long-lived reservoirs in the inter-stellar medium. Accreting gas naturally reproduces the observed HI column density distribution, $\Omega_\text{HI}$, and DLA kinematics, with $\Omega_\text{HI}$ remaining nearly constant despite strong evolution in the cosmic star formation rate. This means that most of the observed HI traces transient, self-shielded accretion flows, of which only a small fraction reaches the ISM to form stars.

Halo-halo mergers, an inherent prediction of a hierarchical universe, are a type of extreme event that can shock-heat diffuse gas and lead to radio continuum emission. Diffusive shock acceleration can generate cosmic-ray electrons along merger
shocks and produce synchrotron radio emission under the influence of the intracluster magnetic field \citep{Kang.2012}. This emission can be observed in the form radio relics, which are thus powerful probes of cluster merger histories and potentially a strong cosmological constraint \citep{Rajpurohit.2021}. Modern large-volume cosmological simulations \citep[e.g.][]{Pillepich.2018,Schaye.2025} have large enough volumes to include many galaxy clusters, while also having sufficient resolution to capture in detail their assembly, internal substructures, and various processes governing their evolution (including halo-halo mergers). This makes such simulations perfectly suited for the study of radio relics,
which form naturally during the simulated cluster mass assembly. Sets of cosmological zoom-in simulations of galaxy clusters can be used for a similar purpose \citep[e.g.][]{Cui.2018,Nelson.2024}. \citet{Lee.2025} used large samples of simulated galaxy clusters to analyse the predicted properties of radio relics in the context of the SKA (see Fig.~\ref{fig:relics} for two examples). They found that even though the brightest relics are associated with massive clusters, the SKA will be sensitive enough to observe mergers between low-mass clusters, which will also be the dominant observed population; see also \citealt{Koribalski01.2026.SKA} for more discussion about the discovery potential of the SKA on the area of radio relics. 
%of double radio relic systems detected with upcoming surveys such as SKA

%\subsection{Testing gastrophysics with radio continuum}

%Probing BH-feedback quenching with radio continuum (Romeel, Nicole, Cedric, Filip)

\section{Next Generation Synthetic Surveys of the Radio Universe}\label{sec:future}

The last decade has seen remarkable progress in simulating the radio sky, as described previously. Yet, a substantial gap remains between current simulations and the ambitious requirements of the SKA. The SKA will conduct all-sky surveys with unprecedented depth, reaching down to dwarf galaxies. Meeting this challenge demands simulations with enormous volumes, exceeding $1\,\rm Gpc^3$, while still resolving galaxies with stellar and gas masses as low as 
$10^6-10^7\,\rm M_{\odot}$.

Achieving this range of scales will continue to require the complementary use of all the approaches described in Section 2 and illustrated in Fig. \ref{fig:scales}. This so-called ``wedding-cake'' strategy combines models of increasing sophistication across decreasing volumes—simple, large-volume frameworks capture the cosmological context, while smaller, higher-resolution simulations provide physical detail. Making this multi-tiered approach successful hinges on developing robust and physically consistent bridges between different tools, resolutions, and simulated volumes.

The key themes that emerge for simulations in the SKA era are the following:

\begin{itemize}
    \item {\it The need for forward modelling.} Direct, one-to-one comparisons between simulations and observations can create misleading tensions if systematic biases affect derived quantities—or, conversely, genuine discrepancies may be hidden when the comparison is not made in the observational frame. Full forward modelling, which transforms simulations into synthetic observables, is therefore essential. Recent work illustrates this clearly: \citet{Marasco.2025} show that simulations struggle to reproduce the quiescent, ordered HI discs seen by MeerKAT; \citet{Chauhan.2021} demonstrate how spectral stacking introduces systematic biases in the H I–halo mass relation; \citet{Brooks.2017,Chauhan.2019,Oman.2022} highlight the need to model the HI emission line to interpret the velocity function; and \citet{Hansen.2024} show that how star-forming and AGN galaxies are separated affects the inferred radio luminosity function. Collectively, these studies underscore the necessity of forward modelling for meaningful simulation–data comparisons. Continued investment in this direction will allow simulations not only to interpret SKA data, but to deliver realistic mock radio skies for survey planning.
    \item {\it Subgrid physics degeneracy.} As shown in Section 3, simulations can produce similar stellar masses and star-formation rates while predicting vastly different gas flows—especially in the distribution and cycling of HI. Comparisons with observations must therefore exploit ensembles of simulations spanning diverse feedback and baryon-cycle prescriptions, enabling identification of which trends are robust and which depend on model assumptions. Within the simulation domain, this diversity is vital for falsifying baryonic models and for isolating the physics driving key observables across cosmic epochs and wavelengths. 
    \item {\it Harnessing new technologies.} Artificial intelligence offers transformative opportunities to accelerate and extend the predictive power of simulations. Neural networks can act as nonlinear emulators, reproducing the results of costly hydrodynamic runs; link N-body and hydrodynamic regimes to bridge scales and volumes; disentangle the effects of feedback parameters on observables; and even map observed quantities directly to underlying physical parameters. The CAMELS project \citep{Villaescusa-Navarro.2021} has pioneered this approach for exploring cosmological constraints under uncertain feedback models. Extending such frameworks to include uncertainties in star formation, radiative transfer, and the mapping between intrinsic and observed properties will help integrate different modelling tools and ensure a smoother, more physically connected implementation of the ``wedding-cake'' strategy. 
    \item {\it Expanding the predictive space.} As SKA capabilities grow, simulations must also broaden their predictive scope to encompass phenomena once considered niche—such as 92 cm deuterium emission, strong and weak lensing, relativistic beaming, jet physics, 21 cm absorption, and Faraday rotation. Modelling these effects will guide SKA observations by identifying the most promising signatures and clarifying the physical insights they can deliver.
\end{itemize}

With the dramatic improvements of the past decade—higher resolution, on-the-fly non-equilibrium thermodynamics down to where 21~cm emission ceases, star formation tied to local collapse conditions rather than H$_2$ abundance, and consistent multi-wavelength galaxy modelling—cosmological simulations have matured from interpretive tools to predictive engines. They are now poised not only to elucidate SKA observations, but also to inform and optimise future SKA survey design and target selection. This shift—from {\it simulations playing a passive interpretive role to becoming active drivers of discovery}—marks a defining opportunity for the theoretical community in the SKA era.

\section{Summary}\label{sec:summary}

The past decade has brought extraordinary advances in modelling the radio sky. Simulations now capture the atomic and molecular gas that fuel star formation, and reproduce the radio continuum emitted by both young stars and active galactic nuclei. These developments have been driven by the rapid improvement of cosmological hydrodynamical simulations and semi-analytic models, which together now provide a coherent, multi-wavelength view of galaxies and their environments. Yet, the capabilities of the Square Kilometre Array (SKA) will far exceed what current models can fully interpret. Its unprecedented sensitivity and sky coverage will probe gas and radio emission down to dwarf galaxies, requiring simulations that combine physical realism with enormous cosmological volume.

Preparing for the SKA era demands a coordinated approach. Large-volume simulations must be linked to high-resolution models through a “wedding-cake” framework, ensuring physical consistency across scales. Forward modelling that helps transform simulations into realistic synthetic observations, will be essential for fair comparison with SKA data. Meanwhile, emerging technologies such as machine learning offer powerful new tools to emulate complex physics, bridge between simulation regimes, and connect theoretical predictions directly to observables. Together, these developments will transform simulations from passive interpreters into active engines guiding SKA survey design, data interpretation, and scientific discovery.

\section*{Acknowledgements}

With the exception of the first author, all other contributing authors are ordered alphabetically. They all contributed with editing, correcting and reading the chapter.

%\subsection{Figures}

%\begin{figure}[h]
%    \centering
%	\includegraphics[width=0.3\columnwidth]{SKAO Pictorial mark-01.png}
%    \caption{Example figure. Place caption below table.}
%    \label{fig:example_figure}
%\end{figure}

%Referring to Fig.~\ref{fig:example_figure}.

%\subsection{Equations}

%\begin{equation}
%    x=\frac{-b\pm\sqrt{b^2-4ac}}{2a}.
%	\label{eq:quadratic}
%\end{equation}
%
%Referring to equation~(\ref{eq:quadratic}).

%\subsection{Tables}

%\begin{table}[h]
%	\centering
%	\caption{Example table. Place caption %above each table.
	%Remember to define the quantities, symbols and units used.}
	%\label{tab:example_table}
	%\begin{tabular}{lccr} % four columns, %alignment for each
	%	\hline
	%	A & B & C & D\\
	%	\hline
%		1 & 2 & 3 & 4\\
%		2 & 4 & 6 & 8\\
%		3 & 5 & 7 & 9\\
%		\hline
%	\end{tabular}
%\end{table}

%Referring to Table~\ref{tab:example_table}.

\bibliographystyle{abbrvnat-maxbibnames4}
\bibliography{chapter} % if your bibtex file is called example.bib

\end{document}